\documentclass[%
 reprint,
superscriptaddress,
 amsmath,amssymb,
 aps,
 prb,
 floatfix,
]{revtex4-2}

\usepackage{amsmath}
\usepackage{array}
\usepackage{dsfont}
\usepackage{mathtools}
\usepackage{graphicx}
\usepackage{dcolumn}
\usepackage{bm}
\usepackage{hyperref}
\usepackage{physics}
\usepackage{empheq}
\usepackage{color}
\usepackage{multirow}
\usepackage{afterpage}
\usepackage{mathrsfs}
\usepackage{tikz}
\usetikzlibrary{shapes}
\usetikzlibrary{tikzmark,calc}

\newcommand{\hexagon}{\mathord{\raisebox{0.6pt}{\tikz{\node[draw,scale=.65,regular polygon, regular polygon sides=6](){};}}}}

\begin{document}

\title{Symmetry fractionalization in the gauge mean-field theory of quantum spin ice}

\author{F\'elix Desrochers}
\email{felix.desrochers@mail.utoronto.ca}
\affiliation{%
 Department of Physics, University of Toronto, Toronto, Ontario M5S 1A7, Canada
}%
\author{Li Ern Chern}%
\affiliation{%
 T.C.M. Group, Cavendish Laboratory, University of Cambridge, Cambridge CB3 0HE, United Kingdom
}%
\author{Yong Baek Kim}%
\email{ybkim@physics.utoronto.ca}
\affiliation{%
 Department of Physics, University of Toronto, Toronto, Ontario M5S 1A7, Canada
}%

\date{\today}

\begin{abstract}
Symmetry fractionalization is a ubiquitous feature of topologically ordered states that can be used to classify different symmetry-enriched topological phases and reveal some of their unique experimental signatures. Despite its vast popularity, there is currently no available framework to study symmetry fractionalization of quantum spin ice (QSI) ---  a $U(1)$ quantum spin liquid (QSL) on the pyrochlore lattice supporting emergent photons --- within the most widely used theoretical framework to describe it, gauge mean-field theory (GMFT). In this work, we provide an extension of GMFT that allows for the classification of space-time symmetry fractionalization. The construction classifies all GMFT \emph{Ans\"atze} that yield physical wave functions invariant under given symmetries and a specific low-energy gauge structure. As an application of the framework, we first show that the only two \emph{Ans\"atze} with emergent $U(1)$ gauge fields that respect all space group symmetries are the well-known 0- and $\pi$-flux states. We then showcase how the framework may describe QSLs beyond the currently known ones by classifying chiral $U(1)$ QSI. We find a new chiral QSL described by $\pi/2$ fluxes of the emergent gauge field threading the hexagonal plaquettes of the pyrochlore lattice. We finally discuss how the different ways translation symmetries fractionalize for all these states lead to unique experimentally relevant signatures and compute their respective inelastic neutron scattering cross-section to illustrate the argument.
\end{abstract}
\maketitle

\section{\label{sec: Introduction} Introduction}
Intrinsic topological phases of matter are novel ground states of many-body systems characterized by long-range entanglement (LRE)~\cite{wen2004quantum, chen2010local, jiang2012identifying, gu2009tensor, levin2005string, levin2006detecting}. LRE leads to drastic phenomenological consequences such as topology-dependent ground state degeneracies and the emergence of deconfined fractional excitations and low-energy gauge structures. The definition of topologically ordered states in terms of LRE is independent of the presence of any symmetries. However, in the presence of symmetries, such as the space group of a lattice or on-site symmetries, topologically ordered phases of matter acquire a finer classification as they can split into different symmetry-enriched topological (SET) classes~\cite{wen2017colloquium, mesaros2013classification, li2017symmetry, lu2012theory, hung2013quantized, teo2015theory}. In distinct SET phases, the global symmetries fractionalize in different ways, i.e., the emergent quasiparticles carry different \emph{fractions}, so to speak, of the local constituents' quantum number (e.g., the charge or spin of the electrons)~\cite{barkeshli2019symmetry,tarantino2016symmetry, tarantino2016symmetry, chen2015anomalous, chen2017symmetry, essin2013classifying, chen2016symmetry, song2015space}. The investigation of symmetry fractionalization in SET classes is a uniquely important tool in our current quest for the experimental realization of topological phases of matter. It provides a classification framework and highlights distinct experimentally accessible signatures since symmetry fractionalization can be measured by conventional shot-noise and neutron scattering experiments~\cite{essin2014spectroscopic, de1998direct, tennant1993unbound, chen2017spectral}. 

Some of the most experimentally relevant potential realizations of topological order are quantum spin liquids (QSLs); quantum paramagnetic ground states of spin systems where competition between different local interactions is so intense that it prevents conventional magnetic long-range order and instead results in LRE~\cite{knolle2019field, savary2016quantum, zhou2017quantum, balents2010spin, broholm2020quantum}. One of the most paradigmatic QSLs is quantum spin ice (QSI). QSI is a QSL on the pyrochlore lattice (see Fig.~\ref{fig:pyrochlore lattice}(a)) with an emergent compact $U(1)$ gauge structure that provides a lattice realization of quantum electrodynamics with a gapless photon-like mode, charged particles with mutual Coulomb interactions (spinons), and magnetic monopoles~\cite{ross2011quantum, benton2012seeing, gingras2014quantum, castelnovo2012spin, chern2019magnetic, udagawa2021spin}. It is known that considering the symmetries of the pyrochlore lattice, QSI can realize at least two different SET phases: the 0- and $\pi$-flux states (0-QSI and $\pi$-QSI) where the hexagonal plaquette of the pyrochlore lattice (see Fig.~\ref{fig:pyrochlore lattice}(c)) are threaded by static 0 and $\pi$ fluxes of the emergent $U(1)$ gauge field respectively~\cite{benton2018quantum, lee2012generic, savary2021quantum, taillefumier2017competing}. Currently, the only available classifications of SET phases on the pyrochlore lattice beyond the 0- and $\pi$-flux states rely on the projective symmetry group (PSG)~\cite{desrochers2022competing, liu2019competing, liu2021symmetric, schneider2022projective}. As introduced by Wen in his seminal work~\cite{wen2002quantum}, the PSG is historically the first attempt to provide a classification scheme for QSLs using space-time symmetry fractionalization. In this framework, a specific parton construction is first assumed. Different PSG classes (i.e., different patterns of space-time symmetry fractionalization) correspond to inequivalent mean-field (MF) solutions within that specific slave-particle construction~\cite{wang2006spin, chern2021theoretical, chern2017quantum, chern2017fermionic, lu2011z, huang2017interplay}. The PSG can classify QSLs invariant under a given set of symmetries, such as fully symmetric QSLs where all space-time symmetries are preserved or chiral QSLs with broken time-reversal symmetry~\cite{bieri2016projective, messio2013time}. It further provides variational wave functions to study the physical properties of these prospective QSLs. For QSI, all PSG classifications have used Abrikosov fermions~\cite{liu2021symmetric, chern2022competing} or Schwinger bosons~\cite{desrochers2022competing, liu2019competing, schneider2022projective} parton constructions. These are generic slave-particle constructions for spin systems that do not have any apparent connection to the physics of QSI, thus making the physical relevance of the identified QSLs dubious.

On the other hand, a parton construction with a transparent connection with QSI is gauge mean-field theory (GMFT)~\cite{savary2012coulombic, savary2013spin, savary2021quantum, lee2012generic}. In this formalism, bosonic spinons hop on the parent diamond lattice while interacting with an emergent compact $U(1)$ gauge field, which directly corresponds with our conceptual understanding of QSI. GMFT is still a widely used theoretical framework to study QSI and has successfully unveiled many vital insights. However, it remains unclear if a classification scheme similar to the PSG can be applied to GMFT. Indeed, there are salient differences between GMFT and other parton constructions upon which the PSG classification is based (i.e., Abrikosov fermions and Schwinger bosons) that make the construction of such a theoretical framework non-trivial. For instance, many ideas from the PSG are challenging to apply to GMFT since the emergent gauge structure, which is the cornerstone of the PSG construction, has an entirely different physical origin. In conventional parton constructions, the emergent gauge field fluctuations are introduced to project back the parton wave function to a physical subspace with a fixed number of partons per site. On the other hand, the emergent gauge structure in GMFT imposes a lattice analog of Gauss's law after artificially introducing a slave bosonic Hilbert space at every site of the parent lattice. Furthermore, the spins in GMFT are represented by directed link variables in contrast to purely on-site operators in the Abrikosov fermions and Schwinger bosons representations. 

In this work, we provide a projective extension of GMFT that allows for the classification of space-time symmetry fractionalization. After briefly reviewing the physics of GMFT, we explain how to find all possible \emph{Ans\"atze} that yield physical wave functions invariant under a specific set of symmetries and construct their corresponding variational MF wave function. With this framework in hand, we first show that assuming the full space group of the lattice, only two QSI states are possible: the 0- and $\pi$-flux states. Even though our extension of GMFT confirms that all fully symmetric QSI states were previously known, it is still an essential step towards the unambiguous experimental realization of QSI since it provides a natural framework that can be extended to study QSLs beyond the fully symmetric $U(1)$ case. For instance, it can be used to classify $\mathbb{Z}_{2}$ QSLs born out of the condensation of spinon pairs or chiral QSLs. The latter classification of chiral QSLs might be especially relevant since recent numerical and analytical studies have found some evidence hinting at the presence of a disordered phase that breaks time-reversal or inversion in proximity to the $SU(2)$ symmetric Heisenberg point~\cite{hering2022dimerization, hagymasi2021possible, astrakhantsev2021broken, burnell2009monopole, kim2008chiral}. Therefore, to exemplify the framework's usefulness, we classify chiral $U(1)$ QSI states. We find two states related by time-reversal symmetry described by $\pi/2$ and $3\pi/2$ fluxes piercing the hexagonal plaquettes. We finally compute the spinon contribution to the neutron scattering cross-section for all these states and show how the spectral periodicity of the two-spinon continuum can be used to distinguish them experimentally.

The rest of the paper is organized as follows: In Sec.~\ref{sec: Conventions} the conventions we use for the pyrochlore and its parent lattice are discussed before reviewing the GMFT construction in Sec.~\ref{sec: Parton Mean-Field Theory}. Our projective extension for GMFT is presented in Sec.~\ref{sec: Projective Classification} and then applied to classify symmetric and chiral $U(1)$ QSI states. We move on to discuss the experimental signatures of these different QSLs and compute the spinons' contribution to their respective neutron scattering cross section in Sec.~\ref{sec: Experimental Signature}. In Sec.~\ref{sec: Discussion}, we finally end with a discussion of our work's implications and future directions.

\section{\label{sec: Conventions} Conventions}

\subsection{\label{subsec: Diamond lattice} Pyrochlore and parent diamond lattice}

\begin{figure}
\includegraphics[width=1.0\linewidth]{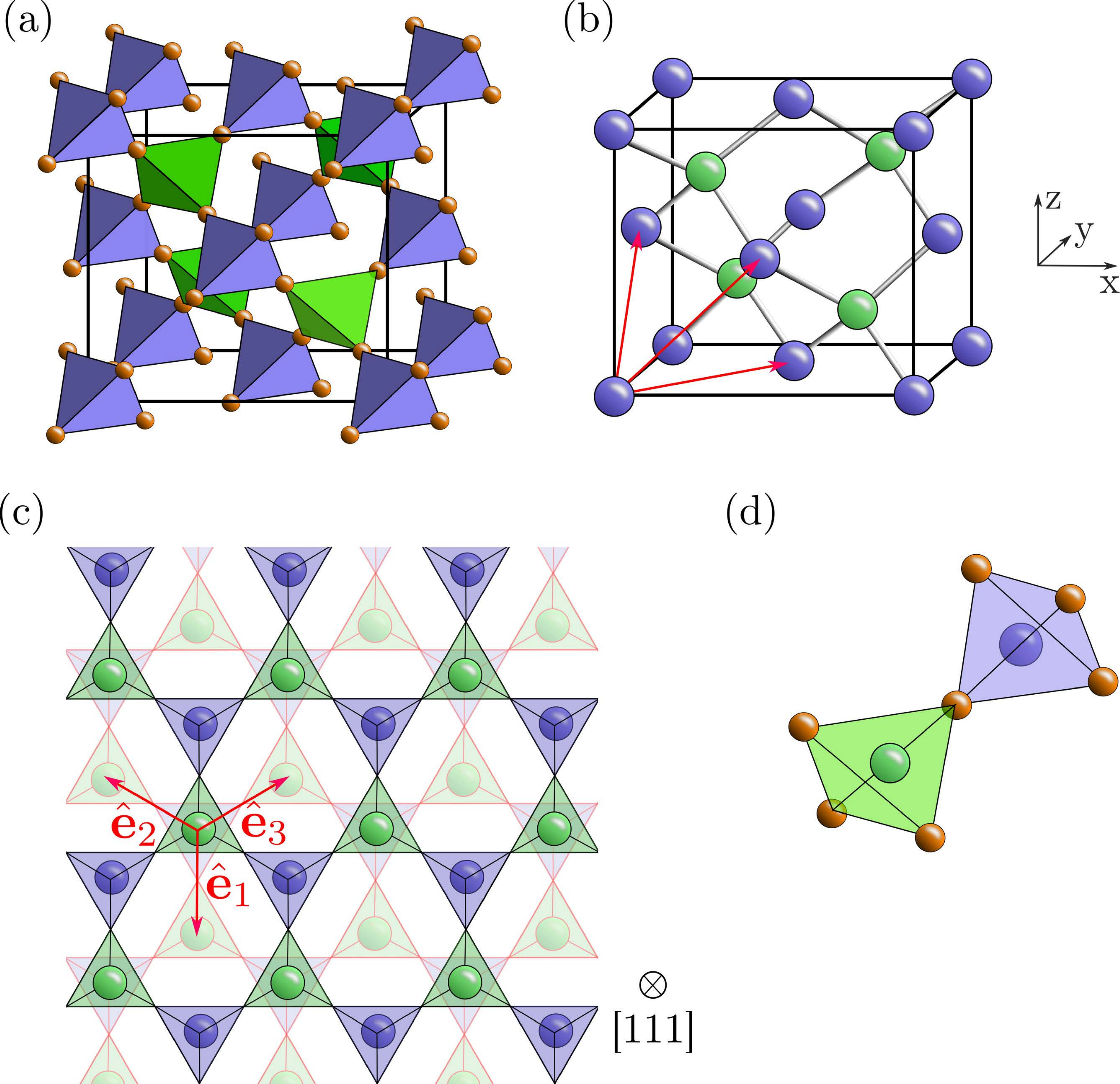}
\caption{ (a) The sites of the pyrochlore lattice form a three-dimensional network of corner-sharing tetrahedra. The down (up) tetrahedra are colored in purple (green). (b) The parent diamond lattice and the three basis translation vectors. (c) When seen along the [111] direction, the pyrochlore lattice forms alternating two-dimensional kagome and triangular layers, whereas the parent diamond lattice forms puckered honeycomb layers with an ABC stacking. (d) The unit cell of the pyrochlore and parent diamond lattice. \label{fig:pyrochlore lattice}}
\end{figure}

The magnetically active ions in spin ice form a pyrochlore lattice, an FCC Bravais lattice with four sublattices shaping into a network of corner-sharing tetrahedra as illustrated in Fig.~\ref{fig:pyrochlore lattice}(a). To identify the position of a unit cell on the pyrochlore lattice, we introduce the \emph{global cartesian coordinates} (GCC), which are the standard frame coordinates of the FCC cube with edge length set to unity, and the following three basis vectors (as expressed in the GCC and illustrated in Fig.~\ref{fig:pyrochlore lattice}(b)-(c))
\begin{subequations} \label{eq: basis for SIPC expressed in GCC}
    \begin{align}
        &\hat{\mathbf{e}}_1 = \frac{1}{2}\left( 0,1,1 \right)\\
        &\hat{\mathbf{e}}_2 = \frac{1}{2}\left( 1,0,1 \right)\\
        &\hat{\mathbf{e}}_3 = \frac{1}{2}\left( 1,1,0 \right).
    \end{align}
\end{subequations}
For later convenience, we also introduce $\hat{\mathbf{e}}_0 = \left( 0,0,0 \right)$. 

The diamond lattice is premedial to the pyrochlore lattice \cite{henley2010coulomb}. It is often colloquially referred to as the dual diamond lattice. To be rigorous, we shall hereafter refer to it as the parent diamond lattice. This parent lattice is an FCC Bravais lattice with two sublattices positioned at the center of the up and down pointing tetrahedra, as shown in Fig.~\ref{fig:pyrochlore lattice}(b). The initial pyrochlore lattice sites are at the center of the bonds on the diamond lattice. Each down tetrahedron (see Fig.~\ref{fig:pyrochlore lattice}(a) for definition) is connected to four nearest-neighbor up tetrahedra by
\begin{subequations} \label{eq: NN vectors in parent diamond lattice}
    \begin{align}
        &\mathbf{b}_0 = \frac{-1}{4}\left( 1, 1, 1 \right)\\
        &\mathbf{b}_1 = \frac{1}{4}\left( -1,1,1 \right)\\
        &\mathbf{b}_2 = \frac{1}{4}\left( 1,-1,1 \right)\\
        &\mathbf{b}_3 = \frac{1}{4}\left( 1,1,-1 \right).
    \end{align}
\end{subequations}
Each up tetrahedron is connected to four down tetrahedra by the opposite vectors. To label the position of the sites on this parent diamond lattice, we introduce the \emph{sublattice indexed diamond coordinates} (SIDC), where the unit cell is identified by a linear combination of the three basis vectors in Eq.~\eqref{eq: basis for SIPC expressed in GCC}. The two sublattices are defined by the sublattice displacement vectors $-\eta_{\alpha} \mathbf{b}_0/2$, where $\eta_{A}=1$ and $\eta_{B}=-1$ with $\alpha$ labeling the sublattice, and $A$ ($B$) stands for down (up). This coordinate system is related to the GCC by 
\begin{align*}
\mathbf{r}_{\alpha} &= \left( r_1, r_2, r_3 \right)_{\alpha} = r_1 \hat{\mathbf{e}}_1 + r_2 \hat{\mathbf{e}}_2 + r_3 \hat{\mathbf{e}}_3 - \frac{\eta_{\alpha}}{2} \mathbf{\mathbf{b}}_{0} \hspace{2.5mm} \text{(SIDC)} \nonumber\\
&= \frac{1}{2} \left( r_2 + r_3, r_1 + r_3, r_1 + r_2 \right) - \frac{\eta_{\alpha}}{2}\mathbf{b}_{0} \hspace{12mm}\text{(GCC)}.  \nonumber
\end{align*}

Finally, spins at every site are defined in a sublattice-dependent \emph{local frame}. The local basis on each pyrochlore sublattice is defined in Appendix~\ref{appendix sec: Local coordinates and generic model}.

\subsection{\label{subsec: Space group} Space group}

The space group (SG) of the diamond lattice is $F d \overline{3} m$ (No. 227). This space group is minimally generated by five operators: three translations $T_{i}$ ($i=1,2,3$), a rotoreflection $\overline{C}_6$  (i.e., $\overline{C}_6 = IC_3$ where $C_3$ is a threefold rotation around $\left[111\right]$ and $I$ is the inversion), and a non-symmorphic screw operation $S$. These space group generators act on the position vector written in the SIDC as
\begin{subequations}
\begin{align}
T_{i}:& \mathbf{r}_{\alpha} \mapsto \left(r_{1}+\delta_{i, 1}, r_{2}+\delta_{i, 2}, r_{3}+\delta_{i, 3}\right)_{\alpha} \\
\overline{C}_{6}:& \mathbf{r}_{\alpha} \mapsto \left(-r_{3}, -r_{1}, -r_{2} \right)_{\pi_{A,B}(\alpha)}\\
S:& \mathbf{r}_{\alpha} \mapsto \left(-r_{1},-r_{2}, r_{1}+r_{2}+r_{3}+\delta_{\alpha,A}\right)_{\pi_{A, B}(\alpha)},
\end{align}
\end{subequations}
where $\pi_{A,B}(\alpha)$  are cyclic permutations of the $A$ and $B$ sublattices.

\section{\label{sec: Parton Mean-Field Theory} Parton Mean-Field Theory}

\subsection{\label{subsec: Slave spinon Formulation} Slave-spinon formulation}

For completeness, we briefly review the physics of quantum spin ice (QSI) and the GMFT parton construction. For a more detailed exposition of the formalism, we refer the reader to Refs.~\cite{savary2012coulombic, savary2013spin, savary2021quantum, lee2012generic}. 

For the sake of simplicity, we restrict our analysis to the XXZ model
\begin{align}
    \mathcal{H}_{\text{XXZ}} = \sum_{\langle \mathbf{R}_{i} \mathbf{R}_{j}' \rangle} \left( J_{zz}  \mathrm{S}_{\mathbf{R}_{i}}^{z} \mathrm{S}_{\mathbf{R}_{j}'}^{z} - J_{\pm} \left( \mathrm{S}_{\mathbf{R}_{i}}^{+} \mathrm{S}_{\mathbf{R}_{j}'}^{-} + \mathrm{S}_{\mathbf{R}_{i}}^{-} \mathrm{S}_{\mathbf{R}_{j}'}^{+} \right) \right),
\end{align}
where the spins are written in the local frame, and the sum is over nearest-neighbor sites of the pyrochlore lattice. We consider the spins to be effective spin-1/2 doublets that transform as usual spinors under elements of the local site symmetry group $D_{3d}$~\cite{rau2019frustrated}. In the antiferromagnetic Ising limit (i.e., $J_{\pm}/J_{zz}\to0$ and $J_{zz}>0$), the $J_{zz}$ coupling enforces the sum over the $z$-component of the spins to be zero for every tetrahedron (i.e., 2-in-2-out). This set of local constraints, known as the ice rules, is a lattice equivalent of the requirement for the spin field to be divergenceless and leads to an extensive ground state degeneracy. The effect of a small transverse term $J_{\pm}$ can then be treated perturbatively within this manifold of 2-in-2-out states. By going to the third order in degenerate perturbation theory, the effective Hamiltonian is a compact $U(1)$ lattice gauge theory of the form~\cite{hermele2004pyrochlore}
\begin{align}
    \mathcal{H}_{\text {eff }} &\sim-J_{\pm}^{3} / J_{z z}^{2} \sum_{\hexagon} \cos (\nabla \times \bar{A}),
\end{align}
where the sum is taken over hexagonal plaquette and the lattice curl $(\nabla \times \bar{A})_{\hexagon} \equiv \sum_{\langle i,j \rangle \in \hexagon} \bar{A}_{i,j}$ is equal to the flux through the hexagonal loops (see Fig.~\ref{fig:pyrochlore lattice}(c)). This emergent $U(1)$ gauge structure is, in a sense, inevitable since the sum over the $z$-component of all spins within any tetrahedra must commute with the effective Hamiltonian as it is defined in a manifold where $\sum_{i \in t} \mathrm{S}_{i}^{z}=0$. For a ferromagnetic transverse coupling $J_\pm>0$, the existence of a deconfined $U(1)$ QSL (i.e., QSI) with 0-flux threading the hexagonal plaquettes (i.e., $(\nabla \times \bar{A})_{\hexagon}=0$) is well established from quantum Monte Carlo (QMC) simulations~\cite{huang2020extended, banerjee2008unusual, huang2018dynamics, shannon2021quantum}. In this perturbative regime, the exact mapping between $J_\pm<0$ and $J_\pm>0$~\cite{savary2021quantum, henley2010coulomb} indicates that a $\pi$-flux QSI state (i.e., $(\nabla \times \bar{A})_{\hexagon}=\pi$) exist for $J_\pm<0$. However, the sign problem of QMC in that parameter regime makes the fate of the $\pi$-flux state ambiguous beyond the perturbative Ising regime.

To go beyond the perturbative regime $|J_{\pm}|\ll J_{zz}$, the theory cannot be restricted to the 2-in-2-out manifold since tetrahedra configurations that do not respect the ice rules then play a significant role. GMFT is a theory introduced by Savary and Balents~\cite{savary2012coulombic} that attempts to properly describe the $U(1)$ deconfined phase without appealing to any perturbative argument. In this framework, bosonic particles that conceptually correspond to defect tetrahedra breaking the ice rules are introduced at the center of each tetrahedron on the parent diamond lattice. The Hilbert of interest is therefore augmented to $\mathscr{H}_{\text{big}} = \mathscr{H}_{\text{spin}} \otimes \mathscr{H}_Q$, where $\mathscr{H}_{\text{spin}}= \otimes_N \mathscr{H}_{S=1/2}$ is the initial Hilbert space of the spins-1/2 on the pyrochlore lattice and $\mathscr{H}_Q$ is the Hilbert space for the new bosonic field $Q_{\mathbf{r}_{\alpha}} \in \mathbb{Z}$ that is defined on each parent diamond lattice site $\mathbf{r}_{\alpha}$. The canonically conjugate variable to the bosonic charge is $\varphi_{\mathbf{r}_{\alpha}}$ (i.e., $\comm{\varphi_{\mathbf{r}_{\alpha}}}{Q_{\mathbf{r}_{\alpha}}}=i$). This naturally leads to the definition of raising and lowering operators $\Phi_{\mathbf{r}_{\alpha}}^\dag = e^{i\varphi_{\mathbf{r}_{\alpha}}}$ and  $\Phi_{\mathbf{r}_{\alpha}} = e^{-i\varphi_{\mathbf{r}_{\alpha}}}$ respectively. To project back $\mathscr{H}_{\text{big}}$ onto the initial physical spin Hilbert space, the discretized Gauss's law
\begin{equation}
    Q_{\mathbf{r}_{\alpha}} = \eta_{\alpha} \sum_{\mu=0}^3 \mathrm{S}^{z}_{\mathbf{r}_{\alpha}+\eta_{\alpha}\mathbf{b}_\mu/2} , \label{eq: physical constraint charge GMFT}
\end{equation}
needs to be enforced for all tetrahedra. All matrix elements are reproduced with the replacements
\begin{subequations} \label{eq: GMFT mapping spins}
\begin{align}
    \mathrm{S}^+_{\mathbf{r}_{A}+ \mathbf{b}_{\mu}/2} &= \Phi^{\dag}_{\mathbf{r}_A} \left( \frac{1}{2} e^{i A_{\mathbf{r}_{\alpha},\mathbf{r}_{\alpha} + \mathbf{b}_\mu}} \right) \Phi_{\mathbf{r}_{A}+\mathbf{b}_\mu}\\
    \mathrm{S}^z_{\mathbf{r}_{A}+\mathbf{b}_\mu/2} &= E_{\mathbf{r}_{A},\mathbf{r}_{A}+\mathbf{b}_\mu}
\end{align}
\end{subequations}
where $A$ and $E$ are canonical conjugate fields that act within the $\mathscr{H}_{\text{spin}}$ subspace of $\mathscr{H}_{\text{big}}$. The local $z$-component of the spin now corresponds to the emergent electric field, and the raising/lowering operators create a pair of spinons on the parent lattice while creating/annihilating an electric field quanta to respect Eq.~\eqref{eq: physical constraint charge GMFT}. 

With those replacements, the XXZ Hamiltonian is 
\begin{align}
\mathcal{H}_{\text{rotor}} &= \frac{J_{z z}}{2} \sum_{\mathbf{r}_{\alpha}} Q_{\mathbf{r}_{\alpha}}^{2} - \frac{J_{\pm}}{4} \sum_{\mathbf{r}_{\alpha}} \sum_{\mu, \nu \ne \mu} \Phi_{\mathbf{r}_{\alpha}+\eta_{\alpha} \mathbf{b}_{\mu}}^{\dagger} \Phi_{\mathbf{r}_{\alpha}+\eta_{\alpha} \mathbf{b}_{\nu}} \nonumber\\
&\hspace{10mm} \times e^{i\eta_{\alpha}(A_{\mathbf{r}_{\alpha}, \mathbf{r}_{\alpha}+\eta_{\alpha} \mathbf{b}_{\nu}}  - A_{\mathbf{r}_{\alpha}, \mathbf{r}_{\alpha}+\eta_{\alpha} \mathbf{b}_{\mu}}  )}. 
\end{align}
The $J_{zz}$ term represents the energetic cost for the existence of spinons while $J_{\pm}$ leads to hopping of the spinons between different tetrahedra of the same type (i.e., up or down) while being coupled to the gauge field. The Hamiltonian has the following $U(1)$ gauge structure 
\begin{equation} \label{eq: gauge transformations}
\left\{\begin{array}{l}
\Phi_{\mathbf{r}_{\alpha}} \rightarrow \Phi_{\mathbf{r}_{\alpha}} e^{i \chi_{\mathbf{r}_{\alpha}}} \\
\left. A_{\mathbf{r}_{\alpha} \mathbf{r}_{\beta}^{\prime}}\rightarrow A_{\mathbf{r}_{\alpha} \mathbf{r}_{\beta}^{\prime}} - \chi_{\mathbf{r}^{\prime}_{\beta}} + \chi_{\mathbf{r}_{\alpha}}\right. 
\end{array}\right.
\end{equation}
as a direct consequence of the physical constraint~\eqref{eq: physical constraint charge GMFT}. This completes the reformulation of the initial nearest neighbor Hamiltonian as a compact $U(1)$ lattice gauge theory coupled to quantum rotors.

\subsection{\label{subsec: Saddle-point approximation} Saddle-point approximation}

The total partition function, taking into account both partons and gauge fields, is
\begin{align}
    \mathcal{Z}=&\int \mathcal{D}[\Phi^{*},\Phi, Q,A,E,\lambda,\zeta] e^{-S_{\text {matter}} - S_{\text {EM}}},
\end{align}
where $S_\mathrm{EM}=\int_{0}^{\beta}\dd{\tau}\frac{U}{2} \sum_{\left\langle\mathbf{r}_\alpha \mathbf{r}_\beta^{\prime}\right\rangle} (E_{\mathbf{r}_{\alpha} \mathbf{r}_{\beta}^{\prime}}^{\tau })^{2}  $ enforces the odd vacuum condition $E_{\mathbf{r}_{\alpha} \mathbf{r}_{\beta}^{\prime}}=\pm 1/2$ by taking the $U\to \infty$ limit, and $S_{\text {matter}}$ describes the quantum rotors coupled to the $U(1)$ gauge field
\begin{align}
    S_{\text{matter}}=&\int_{0}^{\beta} \dd{\tau} \left(\sum_{\mathbf{r}_{\alpha}}\left(i Q_{\mathbf{r}_{\alpha}}^{\tau} \partial_{\tau} \varphi_{\mathbf{r}_{\alpha}}^{\tau}+i \lambda_{\mathbf{r}_\alpha}^{\tau} \left(\Phi_{\mathbf{r}_{\alpha}}^{\tau *} \Phi_{\mathbf{r}_{\alpha}}^{\tau} -1\right)\right. \right. \nonumber \\
    & \left.\left.  + i\zeta_{\mathbf{r}_{\alpha}}^{\tau}\left(\sum_{\mu} E_{\mathbf{r}_\alpha, \mathbf{r}_\alpha + \eta_{\alpha} \mathbf{b}_{\mu}}^{\tau}-Q_{\mathbf{r}_\alpha}^{\tau}\right) \right)  + \mathcal{H}_{\text{rotor}}   
    \right).
\end{align}
The Lagrange multipliers $\lambda_{\mathbf{r}_\alpha}^{\tau}$ and $\zeta_{\mathbf{r}_\alpha}^{\tau}$ enforce the constraint $|\Phi_{\mathbf{r}_{\alpha}}^{\dagger}\Phi_{\mathbf{r}_{\alpha}}|=1$ and Eq.~\eqref{eq: physical constraint charge GMFT} respectively at all sites of the diamond lattice.

To get a tractable model, a saddle point approximation is performed by fixing the gauge field to a constant background (i.e., $A\to \bar{A}$), which amounts to decoupling the dynamics in $\mathscr{H}_{\text{spin}}$ and in $\mathscr{H}_{Q}$. We also allow the gauge charges to take on any integer value $Q_{\mathbf{r}_{\alpha}}\in(-\infty,\infty)$ instead of being constrained to $|Q_{\mathbf{r}_{\alpha}}|<2S$. Doing so and integrating out the charges yields
\begin{align}
    \mathcal{Z}_{\text{MF}}= \int \mathcal{D}[\Phi^{*}, \Phi]  e^{-S_{\text{GMFT}}},
\end{align}
where the saddle point action is
\begin{align} \label{eq: initial GMFT action}
    S_{\text {GMFT}} = \int_{0}^{\beta} \dd{\tau} &\left(  \sum_{\mathbf{r}_{\alpha}} \frac{1}{2 J_{zz}}  \partial_{\tau}\Phi_{\mathbf{r}_\alpha}^{\tau *} \partial_{\tau} \Phi_{\mathbf{r}_\alpha}^{\tau} + \mathcal{H}_{\text {GMFT}} \right. \nonumber \\
    &\quad\left. + i\sum_{\mathbf{r}_{\alpha}} \lambda_{\mathbf{r}_{\alpha}}^{\tau} \left(\Phi_{\mathbf{r}_{\alpha}}^{\tau*} \Phi_{\mathbf{r}_{\alpha}}^{\tau} -1\right) \right)
\end{align}
and $\mathcal{H}_{\text {GMFT}}$ is the $J_{\pm}$ term in $\mathcal{H}_{\text{rotor}}$ but with the gauge fields fixed to constant values. At this stage, an \emph{Ansatz} is usually made about the gauge field background $\bar{A}$. For instance, with ferromagnetic (antiferromagnetic) transverse coupling, one assumes a gauge field configuration with 0-flux ($\pi$-flux) threading the hexagonal plaquettes as a consequence of the perturbative argument outlined above. However, one must wonder if these two \emph{Ans\"atze} are the only possible $U(1)$ QSLs that respect all lattice symmetries and if there is a way to systematically find all gauge field \emph{Ans\"atze} that respect a given set of symmetries. In the following section, we describe such a framework and classify all fully symmetric and chiral $U(1)$ QSLs within the GMFT parton construction. 

Once the gauge background has been specified, the ground state within this parton mean-field theory approach can be identified. The quantum rotor formalism captures many different phases. Most importantly, it can describe $U(1)$ QSLs that formally correspond to deconfined phases within the traditional $U(1)$ lattice gauge theory. From this deconfined phase, transition to an ordered phase is described either by the Higgs mechanism that occurs through condensation of the spinon (i.e., $\expval{\Phi}\ne 0$) or gauge field confinement (i.e., $\expval{e^{iA}}= 0$). The framework also describes $\mathbb{Z}_2$ QSLs born out of spinon pairs condensation (i.e., $\expval{\Phi \Phi}\ne 0$) from the $U(1)$ QSL in the presence of spinon-spinon interactions.

\section{\label{sec: Projective Classification} Projective Classification}

\begin{figure}
\includegraphics[width=0.9\linewidth]{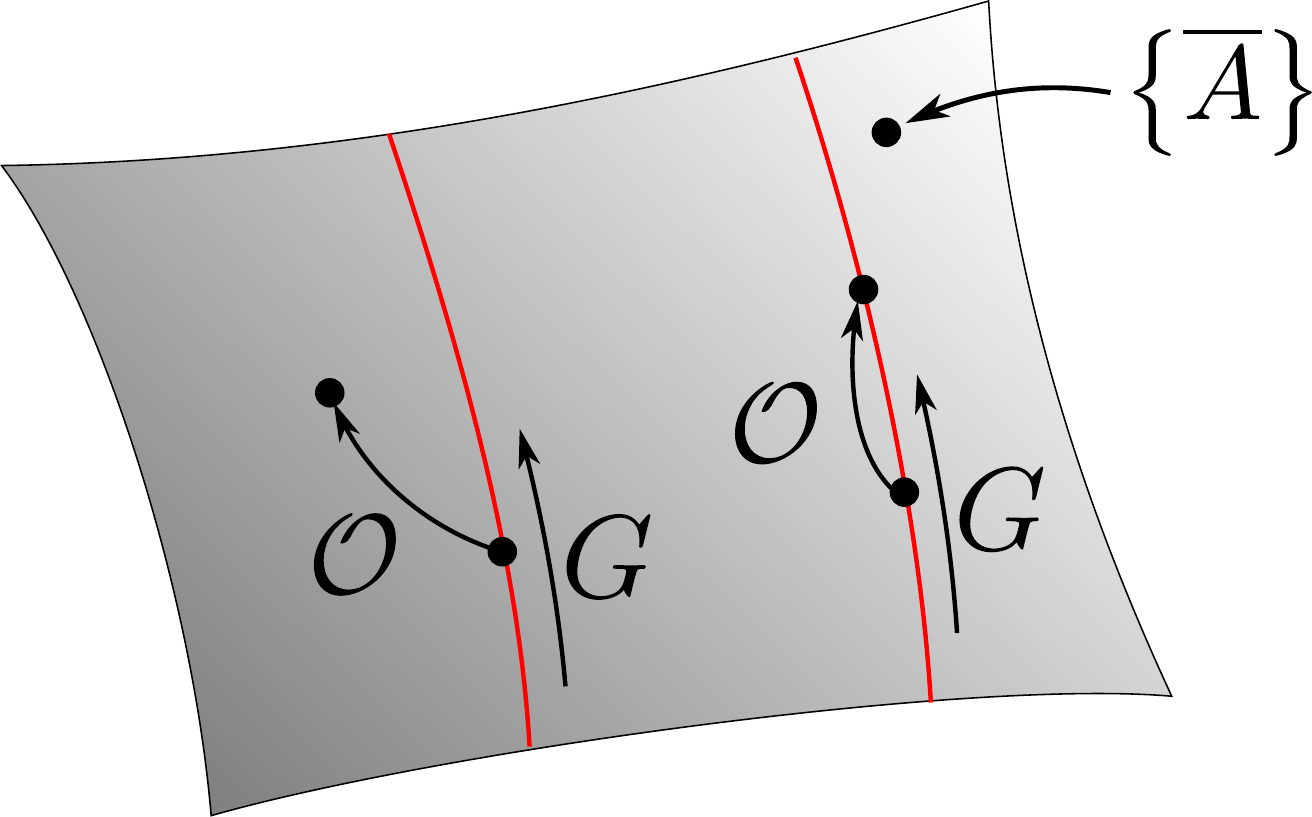}
\caption{Graphical representation of the projective construction. The space represents all possible gauge configurations $\{\overline{A}\}$, and the red curves correspond to sets of gauge configurations related by a gauge transformation $G$ (i.e., different equivalence classes). Depending on the set of equivalent gauge configurations, a symmetry $\mathcal{O}$ can map a representative of an equivalence class to an equivalent (right part of the figure) or non-equivalent (left part of the figure) field configuration. In the first case (right), there exists a gauge transformation $G_{\mathcal{O}}$ such that  $G_{\mathcal{O}}\circ \mathcal{O}$ maps the representative point to itself, whereas no such gauge transformation can be found in the second case (left). The MF eigenstate on the right (left) yields a physical spin wave function that is symmetric (not symmetric) under $\mathcal{O}$. \label{fig: projective symmetry}}
\end{figure}

\subsection{\label{subsec: Generalities} Generalities}

We here discuss the general ideas behind the projective classification of SET phases within the GMFT parton construction. This construction is inspired by the projective symmetry group (PSG) analysis. For a detailed discussion of the PSG, we refer the interested reader to Refs.~\cite{wen2002quantum, bieri2016projective, messio2013time, wang2006spin}. 

After performing a saddle point approximation by replacing the gauge connection operators with a fixed background, the theory does not have a $U(1)$ gauge structure since Gauss's law is not respected anymore. This stems from the decoupling of the gauge $\mathscr{H}_{\text{spin}}$ and bosonic $\mathscr{H}_{Q}$ Hilbert spaces. However, even if the gauge structure is absent at the MF level, it still has important consequences. To see this, one can consider the action of the operator generating the $U(1)$ gauge transformations
\begin{align}
U(\{\chi\})&=\prod_{\mathbf{r}_\alpha} \exp\left(i\chi_{\mathbf{r}_\alpha} \left(Q_{\mathbf{r}_\alpha}- \sum_{\mu} E_{\mathbf{r}_{\alpha},\mathbf{r}_{\alpha}+\eta_{\alpha}\mathbf{b}_\mu}\right)\right)
\end{align}
on a GMFT eigenstate assuming a specific gauge background $\left\{ \overline{A} \right\}$
\begin{equation}
    \mathcal{H}_{\text{GMFT}} \left(\left\{ \overline{A} \right\}\right) \ket{\Psi \left(\left\{ \overline{A}  \right\}\right) } = \mathcal{E}\left(\left\{ \overline{A}  \right\}\right)  \ket{\Psi \left(\left\{ \overline{A} \right\}\right)}.
\end{equation}
This operator maps the gauge configuration to another one $U:\left\{ \overline{A} \right\}\mapsto \left\{ G(\overline{A}) \right\} \equiv \left\{ \overline{A}_G \right\}$ by a transformation of the form expressed in Eq.~\eqref{eq: gauge transformations}. To simplify the notation, we use a subscript to denote the transformed gauge field and suppress the explicit mention of the $\chi$ parameters. It is first straightforward to see that all GMFT eigenstates with gauge configurations related by gauge transformations (i.e, $\ket{\Psi (\{ \overline{A}\}) }$ and $\ket{\Psi (\{ \overline{A}_G\}) }$) are degenerate. Indeed,
\begin{align*}
    &\mathcal{H}_{\text{GMFT}} \left(\left\{ \overline{A}_G \right\}\right)  \ket{\Psi \left(\left\{ \overline{A}_G  \right\}\right) } = \mathcal{E}\left(\left\{ \overline{A}_G  \right\}\right)  \ket{\Psi \left(\left\{ \overline{A}_G \right\}\right)} \nonumber\\
    &=U \mathcal{H}_{\text{GMFT}} \left(\left\{ \overline{A} \right\}\right) U^\dagger U \ket{\Psi \left(\left\{ \overline{A}  \right\}\right) } \\
    &= \mathcal{E}\left(\left\{ \overline{A}  \right\}\right) \ket{\Psi \left(\left\{ \overline{A}_G \right\}\right)}
\end{align*}
implies $\mathcal{E}(\{ \overline{A} \}) = \mathcal{E}(\{ \overline{A}_G \})$.

Next, the GMFT eigenfunctions are not physical spin wave functions in general unless they happen to satisfy Eq.~\eqref{eq: physical constraint charge GMFT}. To recover a physical spin wave function one can think of using a projector-like transformation $\mathcal{P}_{\text{Gauss}}$ that removes any charge configuration that does not respect $|Q_{\mathbf{r}_{\alpha}}|<2S$ and acts on the $\mathscr{H}_{\text{spin}}$ part of $\ket{\Psi (\{ \overline{A}\}) }=\ket{\{E\}}\otimes\ket{\{Q\}}$ such that $\mathcal{P}_{\text{Gauss}}\ket{\Psi (\{ \overline{A}\})}$ respects the constraint of Eq.~\eqref{eq: physical constraint charge GMFT}. Accordingly, since $\comm{U}{\mathcal{P}_{\text{Gauss}}}=0$ because $U$ acts trivially on any state that respects the lattice Gauss's law, all GMFT eigenstates that only differ by a gauge transformation yield the same physical spin wave function
\begin{align*}
    & \mathcal{P}_{\text{Gauss}} U \ket{\Psi (\{ \overline{A}\})} = \mathcal{P}_{\text{Gauss}}\ket{\Psi (\{ \overline{A}_{G}\})}\\
    &= U \mathcal{P}_{\text{Gauss}}  \ket{\Psi (\{ \overline{A}\})} \\
    &= \mathcal{P}_{\text{Gauss}}  \ket{\Psi (\{ \overline{A}\})}.
\end{align*}
This argument is independent of the way one chooses to implement the projection back to the physical spin space $\mathcal{P}_{\text{Gauss}}$. Accordingly, although the gauge structure is not explicitly present, the MF theory still has a redundancy in its description. This redundancy has important and subtle consequences. If we require a GMFT wave function to yield a physical spin state that respects a symmetry $\mathcal{O}$, then this amounts to requiring that $\mathcal{O}$ maps the MF wave function $\ket{\Psi (\{ \overline{A}\})}$ to the same MF state \emph{up to a gauge transformation}. That is, for a GMFT wave function to yield a physical state that is symmetric under a specific transformation, there needs to exist a gauge transformation $G_{\mathcal{O}}$ such that the MF state is invariant under $G_{\mathcal{O}}\circ\mathcal{O}$. Equivalently stated at the Hamiltonian level, a specific gauge background $\{\overline{A}\}$ will yield a symmetric physical state under $\mathcal{O}$ if
\begin{align}
    G_{\mathcal{O}}\circ\mathcal{O}: \mathcal{H}_{\text{GMFT}}\left(\left\{ \overline{A} \right\}\right) \mapsto  \mathcal{H}_{\text{GMFT}}\left(\left\{ \overline{A} \right\}\right).
\end{align}
This idea is illustrated graphically in Fig.~\ref{fig: projective symmetry}. As a result, all static gauge field configurations corresponding to physical states invariant under a given set of symmetries $\{\mathcal{O}_1,\mathcal{O}_2,...\}$ can be classified by identifying the associated gauge-enriched operations $\{\tilde{\mathcal{O}}_1, \tilde{\mathcal{O}}_2,...\} = \{G_{\mathcal{O}_1}\circ\mathcal{O}_1,G_{\mathcal{O}_2}\circ\mathcal{O}_2,...\}$ that can leave $\mathcal{H}_{\text{GMFT}}$ invariant. 

As a final remark, we note that the subgroup of pure gauge transformations $G_{\text{IGG}}\circ\mathds{1}$ that leave the MF Hamiltonian invariant is not associated with any symmetry of the initial system but rather the emergent low-energy gauge structure of the model. This subgroup is called the invariant gauge group (IGG). From an algebraic standpoint, the gauge-enriched group $\tilde{F}$ is a central extension of the original group $F=\{\mathcal{O}_1,\mathcal{O}_2,...\}$ by the IGG ($F = \tilde{F}/\text{IGG}$), and the second cohomology group classifies all inequivalent gauge-enriched classes. 

Looking at Eq.~\eqref{eq: initial GMFT action}, it can be observed that the IGG of the GMFT action is $U(1)\times U(1)$ since the two diamond sublattices are decoupled. This is a very particular property that stems from our restriction to the XXZ model. The IGG would reduce to $U(1)$ by adding any coupling between the two sublattices. Such coupling could be induced by interactions beyond the XXZ model. Since the  $U(1)\times U(1)$ gauge structure is a fine-tuned and fragile property whose naturalness is ambiguous, we will restrict our attention in the rest of the paper to the more physically relevant case where the IGG is $U(1)$. This will also allow us to make contact with existing literature on QSI. This potential $U(1)\times U(1)$ gauge structure remains an interesting observation that requires clarification.


\subsection{\label{subsec: Projective construction} Projective construction}

We now discuss the detailed implementation of the formalism. Within GMFT, the spin operators are mapped to directed link variables. Consequently, it can be remarked that time-reversal and certain space group operations map a bosonic creation operator to an annihilation one and vice versa. For instance, a transformation acting purely on the lattice that changes the orientation of the links (i.e., maps bonds $A\to B$ to $B\to A$) effectively inverts the annihilation and creation operator. As such, the symmetry operations can not be adequately represented by acting solely on $\Phi_{\mathbf{r}_\alpha}$. To build a representation of the space-time symmetry operations on the bosonic fields that correctly encodes this information, we thus introduce the vector field
\begin{align} \label{eq: bosonic vector field}
    \vec{\Psi}_{\mathbf{r}_{\alpha}} = \mqty( \Phi_{\mathbf{r}_{\alpha}}\\ \Phi_{\mathbf{r}_{\alpha}}^{\dagger}),
\end{align}
and the gauge matrices
\begin{align}
    \mathcal{G}_{\overline{A}}(\mathbf{r}_{\alpha},\mathbf{r}_{\alpha}+\eta_{\alpha}\mathbf{b}_\mu) =
    \mqty( e^{i\eta_{\alpha} \overline{A}_{\mathbf{r}_\alpha, \mathbf{r}_\alpha + \eta_{\alpha} \mathbf{b}_{\mu} }} & 0 \\ 
    0 & e^{-i\eta_{\alpha} \overline{A}_{\mathbf{r}_\alpha,\mathbf{r}_\alpha + \eta_{\alpha} \mathbf{b}_{\mu}}} ) .
\end{align}
With this notation, the GMFT Hamiltonian is written as
\begin{widetext}
\begin{align}
    \mathcal{H}_{\text{GMFT}}
     &= - \frac{J_{\pm}}{8} \sum_{\mathbf{r}_{\alpha}} \sum_{\mu, \nu \ne \mu}  \vec{\Psi}_{\mathbf{r}_{\alpha}+\eta_{\alpha} \mathbf{b}_{\mu}}^{\dagger} \mathcal{G}_{\overline{A}}(\mathbf{r}_{\alpha}+\eta_{\alpha}\mathbf{b}_\mu,\mathbf{r}_{\alpha}) \mathcal{G}_{\overline{A}}(\mathbf{r}_{\alpha},\mathbf{r}_{\alpha}+\eta_{\alpha}\mathbf{b}_\nu)
    \vec{\Psi}_{\mathbf{r}_{\alpha}+\eta_{\alpha}\mathbf{r}_{\nu}}.
\end{align}
\end{widetext}
and the action of the symmetry generators on the bosonic field can be represented by (see Appendix~\ref{appendix: Transformation of the parton operators})

\begin{subequations}  \label{eq: Transformation of the parton operators}
    \begin{align}
    T_{i}: &
        \vec{\Psi}_{\mathbf{r}_{\alpha}} \mapsto \mqty( 1 & 0 \\ 0 & 1) \vec{\Psi}_{T_{i}(\mathbf{r}_{\alpha})} = \mathcal{U}_{T_{i}} \vec{\Psi}_{T_{i}(\mathbf{r}_{\alpha})} \\
    \overline{C}_6: &
        \vec{\Psi}_{\mathbf{r}_{\alpha}} \mapsto \mqty( 1 & 0 \\ 0 & 1) \vec{\Psi}_{\overline{C}_{6}(\mathbf{r}_{\alpha})} = \mathcal{U}_{\overline{C}_6} \vec{\Psi}_{\overline{C}_{6}(\mathbf{r}_{\alpha})} \\
    S: &
        \vec{\Psi}_{\mathbf{r}_{\alpha}} \mapsto \mqty( 0 & 1 \\ 1 & 0) \vec{\Psi}_{S(\mathbf{r}_{\alpha})} = \mathcal{U}_{S} \vec{\Psi}_{S(\mathbf{r}_{\alpha})}.
    \end{align}
\end{subequations}

As discussed in the previous subsection, a gauge transformation of the form 
\begin{align}
    G: \vec{\Psi}_{\mathbf{r}_{\alpha}} \mapsto \mqty( e^{i \phi\left(\mathbf{r}_{\alpha}\right)} & 0 \\ 0 & e^{-i \phi\left(\mathbf{r}_{\alpha}\right)}) \Psi_{\mathbf{r}_\alpha} = \mathcal{G}(\mathbf{r}_\alpha) \Psi_{\mathbf{r}_\alpha}
\end{align}
is associated with each symmetry operation to make it projective. The projective transformations that act on the spinons are
\begin{subequations} 
  \begin{align}
    \tilde{T}_{i} : \vec{\Psi}_{\mathbf{r}_{\alpha}} &\mapsto \mathcal{G}_{T_{i}}[T_{i}(\mathbf{r}_\alpha)]^{n_{T_{i}}} \mathcal{U}_{T_{i}} \vec{\Psi}_{T_{i}(\mathbf{r}_{\alpha})} \\
    \tilde{\overline{C}}_6 : \vec{\Psi}_{\mathbf{r}_{\alpha}} &\mapsto \mathcal{G}_{\overline{C}_6}[\overline{C}_6(\mathbf{r}_\alpha)]^{n_{\overline{C}_6}} \mathcal{U}_{\overline{C}_6} \vec{\Psi}_{\overline{C}_{6}(\mathbf{r}_{\alpha})}  \\
    \tilde{S} : \vec{\Psi}_{\mathbf{r}_{\alpha}} &\mapsto \mathcal{G}_{S}[S(\mathbf{r}_\alpha)]^{n_{S}} \mathcal{U}_{S} \vec{\Psi}_{S(\mathbf{r}_{\alpha})},
  \end{align}
\end{subequations}
where the gauge matrices are of the form 
\begin{align}
     \mathcal{G}_{\mathcal{O}}[\mathcal{O}(\mathbf{r}_\alpha)] &=  \mqty( e^{i \phi_{\mathcal{O}}\left[\mathcal{O}\left(\mathbf{r}_{\alpha}\right)\right]} & 0 \\ 0 & e^{-i \phi_{\mathcal{O}}\left[\mathcal{O}\left(\mathbf{r}_{\alpha}\right)\right]})
\end{align}
and the factors $n_{T_{i}}=1$, $n_{\overline{C}_6}=1$, $n_{S}=-1$ are introduced to correctly apply the gauge transformation considering if the creation and annihilation operators have been exchanged.

\subsection{\label{subsec: Symmetric U(1) quantum spin ice}  Symmetric \texorpdfstring{$U(1)$}{U(1)} quantum spin ice states}

We are now in a position to classify QSLs that realize different patterns of space-time symmetry fractionalization. We first classify all QSLs which have an IGG of $U(1)$ (i.e., $\phi_{\mathcal{O}}\in[0,2\pi)$ for all phase factors in the gauge-enriched symmetry operations) and respect all space group symmetries. The detailed classification is presented in Appendix~\ref{appendix: Classification of symmetric $U(1)$ spin liquids}. We find that two different classes associated with the phase factors
\begin{subequations} \label{eq: U(1) PSG classification with TRS and inversion}
\begin{align}
\phi_{T_{1}}\left(\mathbf{r}_{\alpha}\right)=& 0 \\
\phi_{T_{2}}\left(\mathbf{r}_{\alpha}\right)=& n_{1} \pi r_{1} \\
\phi_{T_{3}}\left(\mathbf{r}_{\alpha}\right)=& n_{1} \pi\left(r_{1}+r_{2}\right) \\
\phi_{\bar{C}_{6}}\left(\mathbf{r}_{\alpha}\right) =& n_1 \pi r_1 (r_2 + r_3) \\
\phi_{S}\left(\mathbf{r}_{\alpha}\right) =&  \frac{n_{1} \pi}{2} \left(-r_{1}(r_{1}+1) + r_{2}(r_{2}+1) + 2r_{1}r_{2} \right),
\end{align}
\end{subequations}
where $n_1$ is a parameter that can be either 0 or 1. It is not obvious from these equations, but it is later shown that these two GMFT classes are nothing but the well-known 0- and $\pi$-flux states for $n_{1}=0$ and $n_{1}=1$ respectively. 

\subsection{\label{subsec: Chiral U(1) quantum spin ice}  Chiral \texorpdfstring{$U(1)$}{U(1)} quantum spin ice states}

A relevant extension of the previous classification of symmetric QSLs is the classification of chiral QSLs with an IGG of $U(1)$. Chiral QSLs are classically associated with noncoplanar magnetic order~\cite{bose2022chiral, hickey2017emergence}. In contrast to symmetric QSLs, chiral QSLs break time-reversal symmetry and some lattice symmetries modulo a global spin flip~\cite{messio2013time, schneider2022projective}. Namely, a parity $\epsilon_{\mathcal{O}}$ is associated with every symmetry operation $\mathcal{O}$. The parity is defined to be even $\epsilon_{\mathcal{O}}=1$ if the GMFT \emph{Ansatz} respects the symmetry and odd $\epsilon_{\mathcal{O}}=-1$ if the \emph{Ansatz} only respects it modulo time-reversal. There is a subgroup $\chi_{e}$ of the space group with operations that can only be even. Mathematically, all elements of $\chi_{e}$ are sent to the identity by any morphisms from the SG to $\mathbb{Z}_2$. Careful consideration of the diamond lattice SG algebraic constraints shows that the even subgroup $\chi_e$ is generated by $\left\{ T_{1}, T_{2}, T_{3}, C_{3}, C_{3}^{\prime} \right\}$ where we have introduced $C_{3}^{\prime}=S^{-1}C_{3}S$ (see Appendix~\ref{appendix subsec: chiral PSG solution - even subgroup}). On the contrary, the operators $\overline{C}_6$ and $S$ have an undefined parity. The GMFT \emph{Ans\"atze} are then only required to be invariant under operations of the even subgroup and can thus be enumerated by the same procedure we used for the fully symmetric case but using $\chi_{e}$ instead of the whole SG. The chiral classification should capture all symmetric \emph{Ans\"atze} that have been previously identified since these will correspond to the cases where $\overline{C}_6$ and $S$ have an even parity ($\epsilon_{S}=\epsilon_{\overline{C}_6}=1$). The new \emph{Ans\"atze} that can describe chiral QSLs will be those where there is at least one SG operation with an odd parity $\epsilon_{\mathcal{O}}=-1$. Proceeding as such, we find four different GMFT classes with the phase factors
\begin{subequations} \label{eq: U(1) chiral PSG classification}
\begin{align}
\phi_{T_{1}}\left(\mathbf{r}_{\alpha}\right)=& 0 \\
\phi_{T_{2}}\left(\mathbf{r}_{\alpha}\right)=& \frac{ - n_{1/2} \pi}{2} r_{1} \\
\phi_{T_{3}}\left(\mathbf{r}_{\alpha}\right)=& \frac{n_{1/2} \pi}{2} \left(r_{1} - r_{2}\right) \\
\phi_{C_{3}}\left(\mathbf{r}_{\alpha}\right) =& \frac{n_{1/2} \pi}{2} r_1 \left(r_{3} - r_{2}\right) \\
\phi_{C_{3}^\prime}\left(\mathbf{r}_{\alpha}\right) =&  \frac{- n_{1/2} \pi}{4} \left( r_{2}(2r_{1} + r_{2} - 7) + r_{3} (r_{3} - 1 + 2 \delta_{\alpha,0})  \right)
\end{align}
\end{subequations}
where $n_{1/2}\in\{ 0,1,2,3 \}$. The $n_{1/2}=0$ and $n_{1/2}=2$ \emph{Ans\"atze} correspond to the $n_{1}=0$ and $n_{1}=1$ states from our previous fully symmetric classification respectively. The $n_{1/2}=1$ and $n_{1/2}=3$ GMFT \emph{Ans\"atze} are new chiral states that are related by time-reversal. They correspond to a single chiral QSL. Since this classification encompasses the previous symmetric one, we shall only use and refer to the results of the chiral classification in the rest of the paper.

\subsection{\label{subsec: Gauge fields configurations} Gauge field configuration at the saddle point}

\subsubsection{\label{subsubsec: Relating the gauge field on different bonds} Relating the gauge field on different bonds}

The value of the gauge field on every bond needs to be determined to build the GMFT action for the classified \emph{Ans\"atze}. The transformation properties of the gauge field are first necessary to determine the relation between the values of the gauge field on different bonds of the parent diamond lattice. It can be deduced by using the spinon transformations and requiring that the Hamiltonian is invariant under the gauge-enriched symmetry operations. Indeed, the gauge-enriched operators $\widetilde{\mathcal{O}}$ must be symmetries of the GMFT Hamiltonian (i.e., $\widetilde{\mathcal{O}}:\mathcal{H}_{\text{GMFT}}\mapsto \mathcal{H}_{\text{GMFT}}$). For the GMFT Hamiltonian to be invariant under the projective operations, we have the requirement that
\begin{widetext}
\begin{align} \label{eq: mapping of the gauge fields under SG operations}
    \mathcal{G}_{\overline{A}}(\mathcal{O}(\mathbf{r}_{\alpha}),\mathcal{O}(\mathbf{r}_{\alpha}+ \eta_{\alpha}\mathbf{b}_{\mu})) &= \mathcal{U}_{\mathcal{O}}^{\dagger } [\mathcal{G}_{\mathcal{O}}(\mathcal{O}(\mathbf{r}_{A}))^\dagger]^{n_{\mathcal{O}}} \mathcal{G}_{\overline{A}}(\mathbf{r}_{\alpha},\mathbf{r}_{\alpha}+ \eta_{\alpha}\mathbf{b}_{\mu}) [\mathcal{G}_{\mathcal{O}}(\mathcal{O}(\mathbf{r}_{\alpha}+\eta_{\alpha}\mathbf{b}_{\mu}))]^{n_{\mathcal{O}}} \mathcal{U}_{\mathcal{O}},
\end{align}
\end{widetext}
for space group operations.

\subsubsection{\label{subsubsec: Unit cell} Unit cell}

\begin{figure}[b]
\includegraphics[width=1.0\linewidth]{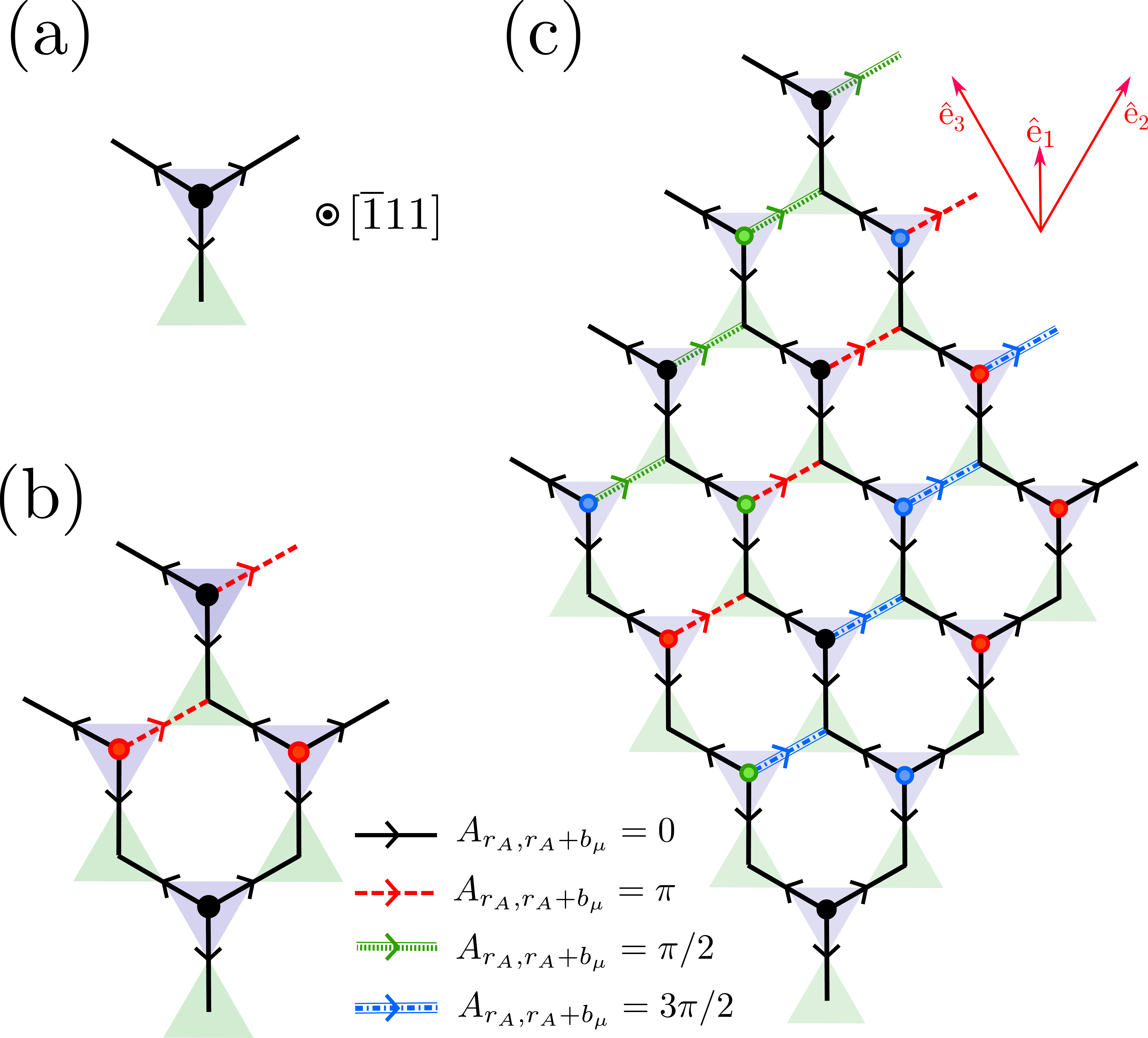}
\caption{Gauge field configuration within the unit cells for the (a) 0-flux ($n_{1/2}=0$), (b) $\pi$-flux ($n_{1/2}=2$) and (c) $\pi/2$-flux ($n_{1/2}=1$) QSI states. As indicated by the arrow, all lines are directed bonds that go from the A (purple) to the B sublattice (green). The full circles represent directed bonds coming out of the plane. \label{fig: Unit cell gauge field configurations}}
\end{figure}

We can use this relation to find the complete gauge field configuration of the $n_{1/2}=$ 0, 1, 2, and 3 GMFT \emph{Ans\"atze}. To do so, the value of the gauge field on a given representative bond is initially arbitrarily fixed. Since all bonds of the parent diamond lattice are related by compositions of the symmetry generators for the symmetric and chiral cases, the value of the gauge field on all other bonds of the lattice can be determined. A translation of the entire GMFT unit cell is a trivial operation. Because $\tilde{T}_{i}=\mathds{1}$ for $n_{1/2}=0$, $\tilde{T}_{i}^2=\mathds{1}$ for $n_{1/2}=2$ and $\tilde{T}_{i}^4=\mathds{1}$ for $n_{1/2}=1$ with $i\in\{2,3\}$, the GMFT unit cell comprises 1, 4 and 16 primitive unit cells of the parent diamond lattice for these three cases respectively. Therefore, the problem of finding the gauge field configuration on all bonds reduces to determining the gauge fields on bonds within a single GMFT unit cell. After proceeding as such (see Appendix~\ref{appendix :  Relation between MF parameters on different bonds in the unit cell}), we find the unit cells depicted in Fig.~\ref{fig: Unit cell gauge field configurations}. The $n_{1/2}=0$ and $n_{1/2}=2$ GMFT \emph{Ans\"atze} are described by static patterns of 0 and $\pi$ gauge field fluxes through all hexagonal loops of the diamond lattice respectively. As a result, we shall refer to them as the  0-flux and $\pi$-flux QSI states in the rest of the paper. We note that these two states are time-reversal invariant even though we did not require it explicitly. The $n_{1/2}=1$ \emph{Ansatz} is described $\pi/2$ fluxes threading the hexagonal plaquette. The flux is equal to the phase spinons acquire after transporting them around a closed loop (i.e. counterclockwise rotation in Fig.~\ref{fig: Unit cell gauge field configurations}). We shall simply refer to this chiral GMFT \emph{Ansatz} as the $\pi/2$-flux state. The $n_{1/2}=3$ \emph{Ansatz} is related to the $n_{1/2}=1$ \emph{Ansatz} by replacing the $\pi/2$ fluxes with $3\pi/2$ fluxes and vice versa (i.e., time-reversal operation). The $\pi/2$- and $3\pi/2$-flux states have the same physical properties. We shall accordingly only consider the $n_{1/2}=1$ state in the rest of this paper. Now that the gauge field configuration has been determined, we can forget about the details of the GMFT classification construction and only retain the definition of the \emph{Ans\"atze} given in Fig.~\ref{fig: Unit cell gauge field configurations}.

\section{\label{sec: Experimental Signature} Experimental Signatures}

With the background gauge field configurations in hand, the GMFT action can be fully constructed and used to evaluate observables for our three prospective QSI states. The only caveat is that the constraint on the rotor length at every site $|\Phi^{\dagger}_{\mathbf{r}_{\alpha}}\Phi_{\mathbf{r}_{\alpha}}|=1$ that is imposed by the site- and time-dependent Lagrange multiplier $\lambda_{\mathbf{r}_{\alpha}}^{\tau}$ is particularly difficult to enforce. To resolve this issue, we follow the usual prescription and perform a large-$N$ approximation by replacing the site-dependent Lagrange multipliers field by sublattice-dependent global ones $\lambda^{\alpha}$ to only enforce the average constraint $\sum_{\mathbf{r}_{\alpha}}\expval{\Phi_{\mathbf{r}_{\alpha}}^{\dagger}\Phi_{\mathbf{r}_{\alpha}}}/N_{\text{d.u.c.}}=\kappa$ for $\alpha\in\{A,B\}$ where $N_{\text{d.u.c.}}$ is the number of diamond lattice primitive unit cell and $\kappa$ is a real parameter. As a side note, we mention that there exist alternatives to this large-$N$ approximation. One of them is the exclusive boson representation of the XY quantum rotor introduced in Ref.~\cite{hao2014bosonic}, which also has the advantage of allowing the straightforward application of standard diagrammatic techniques. Our classification scheme is independent of such a choice. 

After this standard approximation, the translational symmetry of the lattice can be used to Fourier transform our bosonic operators and, in all cases, rewrite the GMFT action in the general form (see Appendix~\ref{appendix: Constructing the MF Hamiltonian})
\begin{align} \label{eq: GMFT action after FT}
    S_{\text{GMFT}} &= \sum_{\mathbf{k},i\omega_n} \sum_{\alpha\in\{A,B\}} \vec{\Phi}^{\dagger}_{\mathbf{k},i\omega_{n},\alpha} \left[\mathscr{G}^{\alpha}(\mathbf{k},i\omega_{n})\right]^{-1} \vec{\Phi}_{\mathbf{k},i\omega_{n},\alpha},
\end{align}
where the wavevector sum is over the reduced first Brillouin zone, the spinon vector field is 
\begin{align} \label{eq: spinon vector field after FT}
    \vec{\Phi}_{\mathbf{k},i\omega_n,\alpha}^{\dagger} = \left( \Phi_{\mathbf{k},i\omega_n,1,\alpha}^{*}, ...,  \Phi_{\mathbf{k},i\omega_n,N_{\text{sl}},\alpha}^{*} \right) 
\end{align}
with the indices labeling all sites of either the $A$ or $B$ sublattices inside the unit cell of a specific GMFT \emph{Ansatz}, and the spinon Matsubara Green's function is
\begin{align} 
    \left[\mathscr{G}^{\alpha}(\mathbf{k},i\omega_{n})\right]^{-1} &= \left( \lambda^{\alpha} + \frac{\omega_n^{2}}{2J_{zz}} \right)\mathds{1}_{N_{\text{sl}}\times N_{\text{sl}}} + M^\alpha(\mathbf{k}).  \label{eq: definition Matsubara Green's function}
\end{align}
$M^{\alpha}(\mathbf{k})$ is an $N_{\text{sl}}\times N_{\text{sl}}$ matrix, with $N_{\text{sl}}$ being the number of primitive diamond lattice unit cells within the unit cell of a specific QSI state. It encodes all information regarding the spinon hopping processes and background gauge field. Identifying the poles of the Green's function (see Appendix~\ref{appendix subsec: Evaluation of observables -> Green's function}), the spinon dispersion is of the form
\begin{align}
    \mathcal{E}^{\alpha}_{\gamma}(\mathbf{k})=\sqrt{2J_{zz}(\lambda^{\alpha}+ \varepsilon_{\gamma}^{\alpha}(\mathbf{k}))},
\end{align}
where $\varepsilon_{\gamma}^{\alpha}(\mathbf{k})$ are the eigenvalues of the $M^{\alpha}(\mathbf{k})$ matrix. Since $N_{\text{sl}}=$ 1, 4, and 16 for the 0-, $\pi$-, and $\pi/2$-flux case respectively, we have two bands for $n_{1/2}=0$ (one band for the $A$ sublattices and one band for the $B$ sublattices), 8 bands for $n_{1/2}=2$, and 32 bands for $n_{1/2}=1$. In all cases, the bands associated with the $A$ sublattices are degenerate with those of the $B$ sublattices. 

\subsection{\label{subsec: An aside on the large-N approximation} An aside on the large-\texorpdfstring{$N$}{N} approximation}

Before discussing specific experimental signatures of the classified QSI states, we make a few comments regarding the large-$N$ approximation. Interpreting the real and imaginary parts of $\Phi_{\mathbf{r}_\alpha}=q_{\mathbf{r}_{\alpha},1}+i q_{\mathbf{r}_{\alpha}, 2}$ as two-dimensional coordinates, the initial hard constraint on the rotor length constrains the system at every site to be on the unit circle $q_{\mathbf{r}_{\alpha},1}^2+q_{\mathbf{r}_{\alpha},2}^2=1$. Such a constraint is important for the mapping between the initial spin model and the slave-particle construction to be exact. The large-$N$ approximation allows the particle to move on the entire two-dimensional plane and only restricts its average displacement. In the existing GMFT literature, $\kappa=1$ is always chosen. However, since the correspondence between the slave-particle construction and the initial spin model is lost by relaxing the hard rotor length constraint, we would like to argue that there are \emph{a priori} no reasons why such a choice ought to be made. Indeed, the $\kappa$ parameter should instead be chosen to reproduce results in a given limit without consideration for the initial hard constraint, in analogy to how the average boson occupancy can be tuned to interpolate between the quantum and classical regime in Schwinger boson mean-field theory~\cite{wang2006spin, sachdev1992kagome, messio2010schwinger}.

In our case of interest, a natural regime where GMFT should be expected to reproduce known results is the Ising or classical spin ice limit (i.e., $J_{\pm}/J_{zz}\to 0$). In such a limit, the spinon dispersion is classically expected to become completely flat at an energy of $J_{zz}/2$~\cite{lacroix2011introduction}. To try and reproduce this result, we first note that in the Ising limit, the rotor length self-consistency equation reduces to (see Appendix~\ref{appendix subsec: Evaluation of observables -> Self-consistency condition})
\begin{align}
    \kappa &= \frac{1}{N_{\text{d.u.c.}}} \sum_{\mathbf{k}} \sum_{\gamma} \frac{J_{zz}}{\mathcal{E}^{\alpha}_{\gamma}(\mathbf{k})} \stackrel{J_{\pm}\to 0}{=} \sqrt{ \frac{J_{zz}}{2 \lambda^{\alpha}}} \Longrightarrow \lambda^{\alpha} = \frac{ J_{zz}}{2\kappa^{2}},
\end{align}
with the corresponding spinon dispersion
\begin{align}
    \mathcal{E}^{\alpha}_{\gamma}(\mathbf{k}) \stackrel{J_{\pm}\to 0}{=} \sqrt{J_{zz}^{2}/ \kappa^{2}}.
\end{align}
In order to respect the classical limit $\mathcal{E}^{\alpha}_{\gamma}(\mathbf{k})=J_{zz}/2$, the parameter $\kappa=2$ needs to be chosen.

\begin{figure*}
    \centering
    \includegraphics[width=1.00\textwidth]{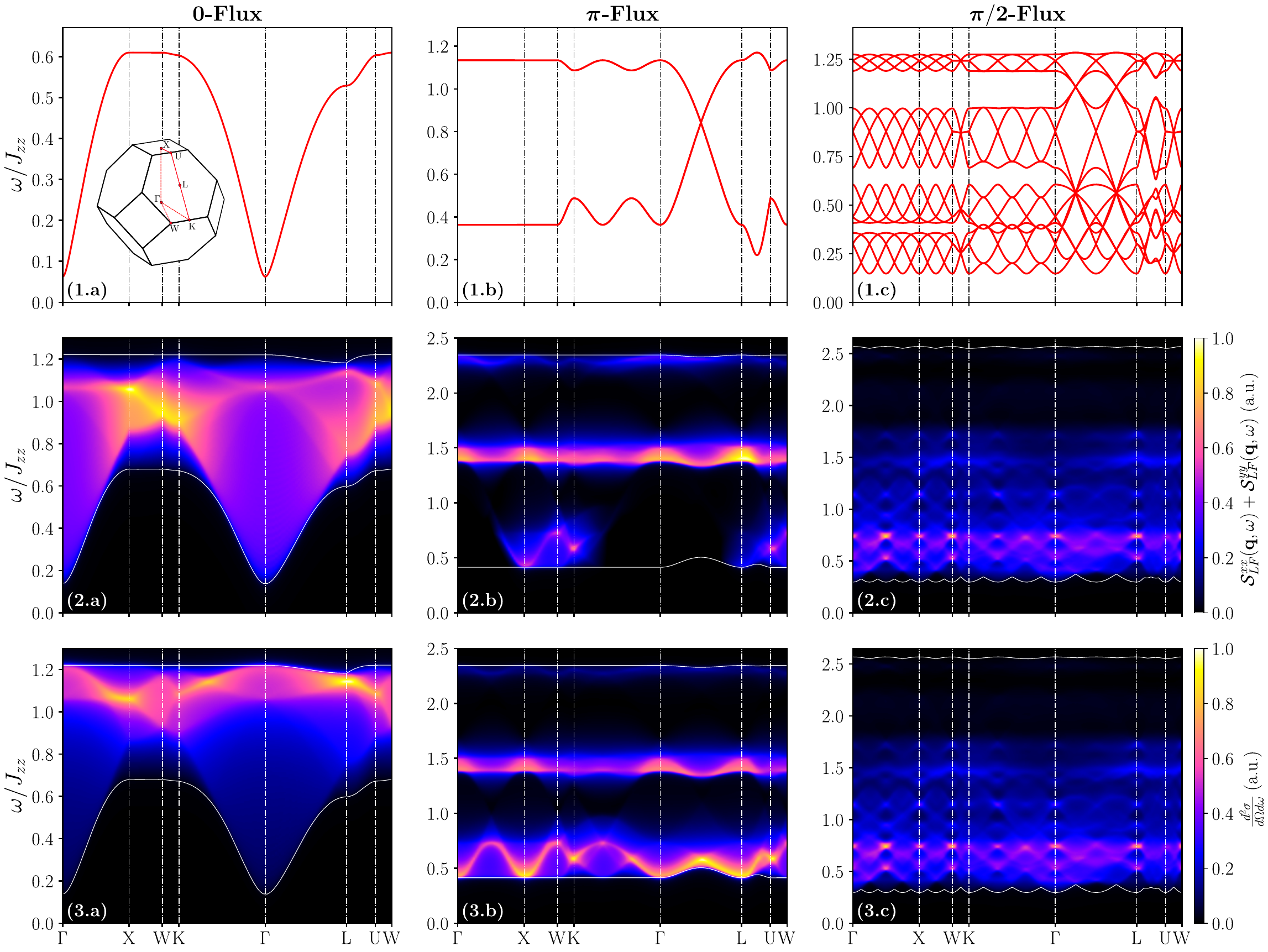}
    \caption{Dynamical signatures of the symmetric and chiral QSI states. (1) Spinon dispersion , (2) dynamical spin correlations in the local sublattice-dependent frame and (3) neutron scattering cross-section for the (a) $0$-flux state at $J_{\pm}/J_{zz}=0.046$, (b) $\pi$-flux state at $J_{\pm}/J_{zz}=-1/3$ and (c)  $\pi/2$-flux state at $J_{\pm}/J_{zz}=-1/3$. Solid white lines denote the upper and lower edges of the two-spinon continuum. The results are broadened by a Lorentzian function with a full width at half maximum of $\eta=0.02$ to mimic finite lifetime effects. The inset of panel (1.a) shows the first Brillouin zone of a face-centered cubic Bravais lattice and its high-symmetry points.}
    \label{fig: DSSF}
\end{figure*}

On top of reproducing the classical spin ice limit, we find that $\kappa=2$ significantly improves the accuracy of the GMFT results. For instance, GMFT with $\kappa=1$ tends to widely overestimate the stability of QSI. For the 0-flux state, GMFT with $\kappa=1$ finds that the QSL is stable until $J_{\pm}/J_{zz}\approx0.192$, whereas the transition to a magnetically ordered phase occurs for the much smaller coupling strength of $J_{\pm}/J_{zz} \approx 0.05$ in QMC simulations~\cite{banerjee2008unusual, shannon2012quantum, kato2015numerical, huang2020extended}. With $\kappa=2$, we find a critical value of $J_{\pm}/J_{zz} \approx 0.048$, which is in surprisingly good agreement with QMC. Using $\kappa=2$, we further find a broad agreement for the position of the lower and upper edges of the two-spinon continuum with QMC results (see Pannels (2.a) and (3.a) of Fig.~\ref{fig: DSSF} compared to results in Ref.~\cite{huang2018dynamics}). For these reasons, all remaining results are presented for $\kappa=2$.

\subsection{\label{subsec: Spectroscopic signatures of space group fractionalization} Spectroscopic signatures of space group fractionalization}

For the $\pi$-flux ($n_{1/2}=2$) and $\pi/2$-flux ($n_{1/2}=1$) states, crystal momentum fractionalizes. Namely, the spinons acquire a non-zero Aharonov-Bohm phase after transporting them around the shortest closed loop since
\begin{align} \label{eq: Aharonov-Bohm phase symmetry}
    \tilde{T}_{i} \tilde{T}_{i+1} \tilde{T}_{i}^{-1} \tilde{T}_{i+1}^{-1} = e^{i \frac{n_{1/2} \pi}{2}},
\end{align}
for $i\in\{1,2,3\}$. Such a fractionalization of translation symmetries has important consequences. It was pointed in Ref.~\cite{essin2014spectroscopic} and later restated for QSI in Ref.~\cite{chen2017spectral}, that such a fractionalization leads to a spectral enhancement of the two-spinon density of states that could be measurable in INS. To see this, we can consider a two-spinon eigenstate with momentum $\mathbf{q}$
\begin{align}
    \ket{\psi} &= \ket{\mathbf{q};z},
\end{align}
where $z$ stands for all other labels like the energy of the state and the spins of the spinons. The momentum is expressed as $\mathbf{q} = q_1\mathbf{G}_{1} + q_2 \mathbf{G}_{2} + q_3 \mathbf{G}_{3}$ where $\mathbf{G}_{i}$ are the reciprocal lattice basis vectors associated with the basis vectors of Eq.~\eqref{eq: basis for SIPC expressed in GCC} (i.e., $\mathbf{G}_i\cdot \hat{\mathbf{e}}_j= 2\pi \delta_{ij}$). Under the assumption of symmetry localization \cite{essin2013classifying}, acting with the translation operator on this two-spinon state amounts to translating the two spinons individually against a translation-invariant background
\begin{align}  \label{eq: translation two-spinon state}
    T_{i}\ket{\psi} = \tilde{T}_{i}(1)\tilde{T}_{i}(2)\ket{\mathbf{q}; z} &= e^{i 2\pi q_{i}}\ket{\mathbf{q}; z},
\end{align}
where $(1)$ and $(2)$ label the spinons, and the last equality follows from the fact that $\ket{\mathbf{q};z}$ is a momentum eigenstate. At the MF level, the spinons are non-interacting. Translating one of them is a good symmetry that leads to eigenstates with the same energy 
\begin{align}
    \tilde{T}_{1}^{m_1}(1) \tilde{T}_{2}^{m_2}(1) \tilde{T}_{3}^{m_3}(1) \ket{\psi} &= \ket{\psi_{(m_1,m_2,m_3)}}.
\end{align}
These degenerate eigenstates may have a different momentum that is not connected by the reciprocal lattice basis vectors, as can be seen from Eq.~\eqref{eq: translation two-spinon state} and repeated usage of Eq.~\eqref{eq: Aharonov-Bohm phase symmetry}
\begin{align}
    & T_{1} \ket{\psi_{ (m_1,m_2,m_3)}} \nonumber \\
    =& \left( \tilde{T}_{1}(1)\tilde{T}_{1}(2) \right) \tilde{T}_{1}^{m_1}(1) \tilde{T}_{2}^{m_2}(1) \tilde{T}_{3}^{m_3}(1) \ket{\psi} \nonumber \\
    =&  e^{i\frac{n_{1/2} \pi}{2} m_{2}} \tilde{T}_{1}^{m_1}(1) \tilde{T}_{2}^{m_2}(1) \tilde{T}_{1}(1)  \tilde{T}_{3}^{m_3}(1) \tilde{T}_{1}(2)  \ket{\psi} \nonumber \\
    =&  e^{i\frac{n_{1/2} \pi}{2} (m_{2}-m_{3})} \tilde{T}_{1}^{m_1}(1) \tilde{T}_{2}^{m_2}(1) \tilde{T}_{3}^{m_3}(1) \tilde{T}_{1}(1) \tilde{T}_{1}(2)  \ket{\psi} \nonumber \\
    =& e^{\left(2\pi i \left( q_i + \frac{n_{1/2}}{4}(m_{2}-m_{3}) \right) \right)} \ket{\psi_{(m_1,m_2,m_3)}}.
\end{align}
Repeating the argument for $T_{2}$ and $T_{3}$, we see that for any state with momentum $\mathbf{q}$ there exist other degenerate eigenstates with the same quantum numbers at momentum $\mathbf{q}+\frac{n_{1/2}}{4}\left(p_1\mathbf{G}_{1} + p_2 \mathbf{G}_{2} + p_3 \mathbf{G}_{3} \right)$ with $p_{i}\in\mathbb{Z}$.

The degeneracy between states connected by fractions of the reciprocal lattice vectors has important consequences. It implies that the two-spinon density of states and the edges of the two-spinon continuum all have to obey 
\begin{align} \label{eq: spectral enhancement}
    \Omega_{n_{1/2}}\left(\mathbf{q}\right) = \Omega_{n_{1/2}}\left(\mathbf{q}+ \frac{n_{1/2}}{4}\left(p_1\mathbf{G}_{1} + p_2 \mathbf{G}_{2} + p_3 \mathbf{G}_{3} \right)\right),
\end{align}
where with $p_{i}\in\mathbb{Z}$. As a specific example, we can consider the path in the reciprocal lattice connecting the $\Gamma$ point to $\text{X}=(\mathbf{G}_{1} + \mathbf{G}_{2})/2$.  For the $\pi$-flux state the spinon dispersion must satisfy $\mathcal{E}_{\gamma}^{\alpha}(\Gamma)=\mathcal{E}_{\gamma}^{\alpha}(\text{X})$ which leads to $\Omega_{n_{1/2}=2}\left(\Gamma\right) = \Omega_{n_{1/2}=2}\left(\text{X}\right)$ for the two-spinon spectrum,
whereas we have $\mathcal{E}_{\gamma}^{\alpha}(\Gamma)=\mathcal{E}_{\gamma}^{\alpha}(\text{X}/2)=\mathcal{E}_{\gamma}^{\alpha}(\text{X})$ and $\Omega_{n_{1/2}=1}\left(\Gamma\right) = \Omega_{n_{1/2}=1}\left(\text{X}/2\right) = \Omega_{n_{1/2}=1}\left(\text{X}\right)$ for the $\pi/2$-flux state. By comparison, there is no constraint between $\Omega_{n_{1/2}=0}\left(\Gamma\right)$ and $\Omega_{n_{1/2}=0}\left(\text{X}\right)$ for the 0-flux state. Similarly, the two-spinon spectrum between the $\Gamma$ and $\text{L}=(\mathbf{G}_{1} + \mathbf{G}_{2} + \mathbf{G}_{3})/2$ points respects $\Omega_{n_{1/2}=2}\left(\Gamma\right) = \Omega_{n_{1/2}=2}\left(\text{L}\right)$ and 
$\Omega_{n_{1/2}=1}\left(\Gamma\right) = \Omega_{n_{1/2}=1}\left(\text{L}/2\right) = \Omega_{n_{1/2}=1}\left(\text{L}\right)$ for the $\pi$- and $\pi/2$-flux states respectively. 

It should be noted that since neutrons couple to spin-1 excitations and the spinons carry spin-1/2, a neutron scattering event corresponds to the creation of a spinon pair. For the deconfined QSI phases, INS probes the two-spinon continuum. The enhanced spectral periodicity described in Eq.~\eqref{eq: spectral enhancement} may be measured in INS and by other spectroscopic probes, thereby potentially offering a practical and accessible way to distinguish between different QSI states experimentally and numerically. 

\subsection{\label{subsec: Dynamical spin structure factor} Dynamical spin structure factor}
Even though the spinon continuum for different QSI states has to respect certain symmetries constraints, it does not necessarily imply that the space group fractionalization will lead to experimentally measurable signatures. Indeed, the inelastic neutron scattering (INS) intensity depends on other factors which do not have to respect the enhanced spectral periodicity. To see if spectral periodicity offers a useful way to distinguish the different QSI states experimentally, we explicitly compute the INS cross-section. The dynamical spin-spin correlations in the local frame are captured by the components of
\begin{align}
\mathcal{S}^{a b}_{LF}(\mathbf{q},\omega)=&\frac{1}{N_{\text{u.c. }}} \sum_{\mathbf{R}_{i}, \mathbf{R}_{j}'} e^{i \mathbf{q} \cdot\left(\mathbf{R}_{i} - \mathbf{R}_{j}'\right)} \nonumber \\
&\hspace{1cm} \times\int \dd{t} e^{i \omega t}  \left\langle \mathrm{S}_{\mathbf{R}_{i}}^a(t) \mathrm{S}_{\mathbf{R}_{j}'}^b(0)\right\rangle ,
\end{align}
where the sum is taken over all sites of the pyrochlore lattice $\mathbf{R}_{i}$, with $i\in\{0,1,2,3\}$ labeling the sublattices. To make direct contact with experimental measurements, the spins need to be rotated from their sublattice-dependent local frame to the global frame, and the coupling between the neutron and the magnetic dipole of the spins is considered by introducing a transverse projector and $g$-factors. The INS cross-section is (neglecting any form factor) proportional to
\begin{align} \label{eq: neutron scattering cross-section}
    &\frac{d^2 \sigma}{d \Omega d \omega} \propto \sum_{a, b}\left(\delta_{ab}-\frac{\mathbf{q}_a \mathbf{q}_b}{|\mathbf{q}|^2}\right) \int \dd{t} e^{i \omega t} \left\langle m^a(\mathbf{q}, t) m^b(-\mathbf{q}, 0)\right\rangle
\end{align}
with
\begin{align}
    &m^a(\mathbf{q}, t) = \frac{1}{\sqrt{N_{\text{u.c.}}}} \sum_{\mathbf{R}_{i}} e^{i \mathbf{q} \cdot \mathbf{R}_{i}} \sum_{c,d} R_\mu^{a c} g^{cd} \mathrm{S}_{\mathbf{R}_{i}}^{d}(t),
\end{align}
where the $R_{\mu}$ matrices are sublattice-dependent rotations from the local frame to the GCC, and the $g$-matrix contains the $g$-factors of the spin in the local frame. 

The $z$-components of the local dynamical correlations $\mathcal{S}^{zz}_{LF}\propto \expval{E E}$ are associated with the emergent photon propagator. One can compute the contribution of the photon to the INS cross-section~\cite{benton2012seeing}. However, since we are not considering gauge fluctuations within our GMFT approach, obtaining a quantitative comparison between the contribution of the matter and the emergent gauge bosons might be challenging. Furthermore, the photons and spinons usually operate on entirely separate energy scales. In most cases, the photon contributes on an energy scale which may be very challenging to resolve experimentally~\cite{huang2018dynamics}. For these reasons, we set $g^{xx}=g^{yy}\ne0$ and $g^{zz}=0$ to consider only the most significant contribution to the dynamical spin structure factor: spinons scattering.

The spinon dispersions, dynamical spin correlations in the local frame, and neutron scattering cross-sections for the three QSI states of interest in a regime where they are not condensed are shown in Fig.~\ref{fig: DSSF}. For the 0-flux state, the INS cross-section shows a broad continuum with most of its spectral weight close to the upper two-spinon continuum edge and high-intensity peaks at the X and L points. It should further be noted that the GMFT calculations are consistent with the QMC results of Ref.~\cite{huang2018dynamics} (see Appendix~\ref{appendix: Comparison of the dynamical spin structure factor for the 0-flux state with QMC} for a detailed comparison). For the $\pi$-flux state, we get bands that are very flat in most directions. This leads to an INS cross-section that is separated into three different energy sectors, where the lowest, central, and highest contributions correspond to processes involving two spinons of the lowest band,  the two different bands, and the upper band respectively. Much of the spectral weight is concentrated in the lower edge of the two-spinon continuum. This clearly makes the spectral enhancement observable especially for the paths $\Gamma\to\text{X}$ and $\Gamma\to\text{L}$. Finally, the 32 bands of the $\pi/2$-flux state lead to a blurry INS spectrum with a few high-intensity peaks. The spectral enhancement might not be as obvious as in the $\pi$-flux case, but we observe a twofold repetition of high-intensity peaks at the same energy for the $\Gamma\to\text{X}$ and $\Gamma\to\text{L}$ paths. 

\section{\label{sec: Discussion} Discussion and future directions}

In this paper, we provided an extension of GMFT to classify all GMFT \emph{Ans\"atze} corresponding to physical spin wave function respecting a given set of symmetries. In the case where the physical spin state resulting from the variational wave function is a deconfined QSL, the theoretical structure classifies SET phases captured by the GMFT parton construction. We explained how subtleties that made the application of ideas from the PSG to GMFT challenging like the origin of the gauge structure and the mapping of the spin operators to directed bond variables could be properly taken into account. 

Application of the theoretical construction shows that there only exist two GMFT \emph{Ans\"atze} that respect all lattice symmetries, the 0- and $\pi$-flux states. There is another chiral QSL that breaks time-reversal symmetry and is described by either $\pi/2$ or $3\pi/2$ fluxes. These states are distinguished by the way translation symmetries fractionalize. We showed how this symmetry fractionalization leads to distinct experimental signatures through the spectral enhancement of the two-spinon density of states. To explicitly confirm that such a signature should be experimentally accessible, we computed the spinon contribution to the INS cross-section and confirmed that the 0-flux state shows no spectral enhancement, whereas the doubling of the unit cell for the $\pi$-flux state is visible. The fourfold spectral enhancement for the chiral $\pi/2$-flux state is also present but might be harder to detect. 

The GMFT approach relies on a saddle point approximation where the gauge field is fixed. One might legitimately question the relevance of our classification scheme considering the disregard of any gauge field fluctuations, especially considering recent indications that the gauge-matter coupling in QSI is stronger than in ordinary QED \cite{pace2021emergent}. Strong gauge interaction in QSI is a topic of great interest that will undoubtedly lead to significant phenomenological consequences. For instance, it was recently shown that the gauge degrees of freedom and their fluctuations are important to understand qualitative and quantitative features in the dynamical correlations of quantum spin ice~\cite{Wilczek2020Spinons, udagawa2021spin}. However, we would like to point out two crucial observations as to why GMFT still provides an essential tool for studying QSI. First, symmetry fractionalization is believed to be a robust characteristic of a topologically ordered phase that is stable to perturbations~\cite{wen2002quantum, barkeshli2019symmetry}. Accordingly, our classification scheme and observations regarding the enhanced spectral periodicity should remain valid in the presence of strong interactions. Second, the excellent agreement between the dynamical spin correlations obtained in QMC and GMFT presented in Appendix~\ref{appendix: Comparison of the dynamical spin structure factor for the 0-flux state with QMC} should serve as a convincing piece of evidence that the correlations obtained in GMFT capture the most important features and can be meaningfully used to compare with experimental results.

Our investigation opens the door to many relevant theoretical investigations. First, the framework we introduced can be used to classify $\mathbb{Z}_{2}$ QSLs. This classification is especially important since the presence of a $\mathbb{Z}_2$ QSL with ferromagnetic transverse coupling was recently suggested by QMC simulations~\cite{huang2020extended}. However, besides its position in the phase diagram, nothing is known about this prospective $\mathbb{Z}_2$ QSL. Furthermore, another recent investigation provided an underlying mechanism to generate interactions between fractionalized quasiparticles coming from the constraint on the physical Hilbert space that could lead to the formation of an intermediate $\mathbb{Z}_2$ QSL between the deconfined $U(1)$ QSL and confining magnetically ordered phase just as observed in QMC \cite{yang2021hidden}. With the numerical hints for the existence of the phase and a potential underlying mechanism to explain its origin, our extension of GMFT could then serve to explore the nature of this intermediate phase. It could be used to classify possible phases, examine their stability, and compute their experimental signatures to compare with possible future QMC results. An interesting question of great current experimental interest is if a similar scenario, the presence of an intermediate $\mathbb{Z}_2$ QSL between the $U(1)$ deconfined and confining phase, is also realized for antiferromagnetic transverse couplings. Such a scenario is much harder to investigate since QMC is plagued by a sign problem in that region of the phase diagram. Still, we hope that by providing the variational wave function of prospective states, our work might help shed some light on this issue. 

Our classification scheme can also be applied for less symmetric variants like the pyrochlore lattice with the application of an electric field~\cite{Hurtubise2017Electric} or the breathing pyrochlore lattice~\cite{benton2015ground, savary2016quantum, tsunetsugu2017theory, essafi2017flat, ezawa2018higher, aoyama2019spin, chern2022competing}. There has been a revival of interest in breathing pyrochlore magnets due to proposals that they may stabilize QSLs with a rank-2 $U(1)$ tensor gauge structure and fractonic excitations~\cite{yan2020rank, han2022realization, zhang2022dynamical}. Many candidate pyrochlore materials have a breathing anisotropy~\cite{tanaka2014novel, okamoto2015magnetic, nilsen2015complex, lee2016multistage, kimura2014experimental, haku2016low, rau2018behavior}. It would be interesting to see how this breaking of the inversion symmetry could lead to potentially new GMFT classes and if a framework analogous to GMFT could be introduced for tensor gauge structures. 

Another exciting direction to take our construction is to apply it to the dipolar-octupolar case~\cite{huang2014quantum, benton2020ground, rau2019frustrated, patri2020theory}. There has been a tremendous interest in this case since the analyses of available data suggest that the pyrochlore compound Ce$_2$Zr$_2$O$_7$ is in a region of parameter space that is believed to stabilize the $\pi$-flux $U(1)$ octupolar QSI~\cite{gao2019experimental, bhardwaj2022sleuthing, smith2022case,gaudet2019quantum, hosoi2022uncovering}. However, even if the position of Ce$_2$Zr$_2$O$_7$ is well known in parameter space, there are still doubts regarding the nature its the ground state. Indeed, the compound is far from the perturbative Ising limit, where the theoretical prediction for the $\pi$-flux $U(1)$ octupolar QSI ground state is well established. It was further recently shown that many experimentally observed key features could be explained by an entirely different $\mathbb{Z}_2$ QSL state with bosonic excitations~\cite{desrochers2022competing}. It would then be interesting to apply our framework and compute the INS cross-section of the octupolar $\pi$-flux QSI state to see how it compares to measurements on Ce$_2$Zr$_2$O$_7$. 

A question that requires further investigation is the naturalness of the chiral $\pi/2$-flux state. We have discussed its properties but have not addressed its stability and potential material realization. It could be stable, especially if one considers coupling constants beyond the XXZ model. A promising regime to investigate would be any path in the parameter space of all possible couplings that interpolates between the classical spin ice Ising limit and a noncoplanar magnetically ordered state. 

Finally, we believe our work might stimulate the development of a framework that could compare SET phases classified in different three-dimensional parton constructions. More broadly, we hope it might provide insights into the study of symmetry fractionalization for three-dimensional topologically ordered phases, a subject still in its infancy.

\begin{acknowledgments}
We thank Kristian Tyn Kai Chung for stimulating discussions. We acknowledge support from the Natural Sciences and Engineering Research Council of Canada (NSERC) and the Centre of Quantum Materials at the University of Toronto. Computations were performed on the Niagara cluster, which SciNet host in partnership with Compute Canada. YBK is also supported by the Guggenheim Fellowship from the John Simon Guggenheim Memorial Foundation and the Simons Fellowship from the Simons Foundation. Some parts of this work were performed at the Aspen Center for Physics, which is supported by the National Science Foundation Grant No. PHY-1607611.
\end{acknowledgments}

\appendix

\section{\label{appendix sec: Local coordinates and generic model} Local coordinates}

There are four sites of the pyrochlore lattice within a primitive unit cell. Their position can be expressed by defining $\hat{\epsilon}_i = \frac{1}{2}\hat{\mathbf{e}}_i$ ($i=1,2,3$) to be the displacement of the $i=1,2,3$ sublattices from the $i=0$ sublattice respectively (where $\hat{\epsilon}_0=\hat{\mathbf{e}}_0=0$). The basis vectors of the local frame at each of these sublattice sites are defined in table~\ref{tab: Local basis}.

\begin{table}[htbp]
\caption{\label{tab: Local basis}%
Local sublattice basis vectors
}
\begin{ruledtabular}
\begin{tabular}{ccccc}
$\mu$ & 0 & 1  & 2  & 3 \\
\hline
$\hat{\mathbf{z}}_{\mu}$ & $\frac{1}{\sqrt{3}}\left(1,1,1\right)$ & $\frac{-1}{\sqrt{3}}\left(-1,1,1\right)$  & $\frac{-1}{\sqrt{3}}\left(1,-1,1\right)$  & $\frac{-1}{\sqrt{3}}\left(1,1,-1\right)$   \\[2mm]
$\hat{\mathbf{y}}_{\mu}$ & $\frac{1}{\sqrt{2}}\left(0,-1,1\right)$  & $\frac{1}{\sqrt{2}}\left(0,1,-1\right)$  & $\frac{-1}{\sqrt{2}}\left(0,1,1\right)$ & $\frac{1}{\sqrt{2}}\left(0,1,1\right)$  \\[2mm]
$\hat{\mathbf{x}}_{\mu}$ & $\frac{1}{\sqrt{6}}\left(-2,1,1\right)$ & $\frac{-1}{\sqrt{6}}\left(2,1,1\right)$  & $\frac{1}{\sqrt{6}}\left(2,1,-1\right)$  & $\frac{1}{\sqrt{6}}\left(2,-1,1\right)$    \\
\end{tabular}
\end{ruledtabular}
\end{table}

\section{\label{appendix: Transformation of the parton operators} Transformation of the parton operators}

We are considering effective spin-1/2 Kramers doublet \cite{rau2019frustrated}. Under the generators of the space group, the pseudos-spins transform as
\begin{subequations} 
  \begin{align}
     T_i :  \qty\Big{ \mathrm{S}^{+}_{\mathbf{R}_{i}}, \mathrm{S}^{-}_{\mathbf{R}_{i}}, \mathrm{S}^{z}_{\mathbf{R}_{i}} } \mapsto &  \qty\Big{  \mathrm{S}^{+}_{T_i(\mathbf{R}_{i})}, \mathrm{S}^{-}_{T_i(\mathbf{R}_{i})}, \mathrm{S}^{z}_{T_i(\mathbf{R}_{i})} }\\
    \overline{C}_6 :  \qty\Big{  \mathrm{S}^{+}_{\mathbf{R}_{i}}, \mathrm{S}^{-}_{\mathbf{R}_{i}}, \mathrm{S}^{z}_{\mathbf{R}_{i}} } \mapsto &  \qty\Big{  \gamma \mathrm{S}^{+}_{\overline{C}_6(\mathbf{R}_{i})},  \overline{\gamma}  \mathrm{S}^{-}_{\overline{C}_6(\mathbf{R}_{i})}, \mathrm{S}^z_{\overline{C}_6(\mathbf{R}_{i})} } \\
    S :  \qty\Big{ \mathrm{S}^{+}_{\mathbf{R}_{i}}, \mathrm{S}^{-}_{\mathbf{R}_{i}}, \mathrm{S}^{z}_{\mathbf{R}_{i}} } \mapsto &  \qty\Big{  -\gamma \mathrm{S}^{-}_{S(\mathbf{R}_{i})} , -\overline{\gamma}  \mathrm{S}^{+}_{S(\mathbf{R}_{i})} , -\mathrm{S}^{z}_{S(\mathbf{R}_{i})} },
  \end{align}
\end{subequations}
where $\gamma=e^{2\pi i/3}$. In terms of the GMFT parton construction, this corresponds to
\begin{widetext}
\begin{subequations} 
  \begin{align}
    T_{i} :&  \qty\Big{ \frac{1}{2} \Phi^{\dag}_{\mathbf{r}_A} e^{i A_{\mathbf{r}_{A},\mathbf{r}_{A}+ \mathbf{b}_\mu}} \Phi_{\mathbf{r}_{A}+\mathbf{b}_\mu},
    \frac{1}{2} \Phi^{\dag}_{\mathbf{r}_A+\mathbf{b}_\mu} e^{-iA_{\mathbf{r}_{A},\mathbf{r}_{A}+ \mathbf{b}_\mu}}  \Phi_{\mathbf{r}_{A}},
    E_{\mathbf{r}_{A},\mathbf{r}_{A}+\mathbf{b}_\mu} }\nonumber \\ 
    &\mapsto  \qty\Big{ \frac{1}{2} \Phi^{\dag}_{T_{i} (\mathbf{r}_A)} e^{i A_{T_{i} (\mathbf{r}_{A}),T_{i} (\mathbf{r}_{A}+ \mathbf{b}_\mu)}}  \Phi_{T_{i} (\mathbf{r}_{A}+\mathbf{b}_\mu)}, \frac{1}{2}
    \Phi^{\dag}_{T_{i} (\mathbf{r}_A+\mathbf{b}_\mu)} e^{-iA_{T_{i} (\mathbf{r}_{A}),T_{i} (\mathbf{r}_{A}+ \mathbf{b}_\mu)}}  \Phi_{T_{i} (\mathbf{r}_{A})},
    E_{T_{i} (\mathbf{r}_{A}),T_{i} (\mathbf{r}_{A}+\mathbf{b}_\mu)} 
    } \\ 
    \overline{C}_{6} :&  \qty\Big{ \frac{1}{2} \Phi^{\dag}_{\mathbf{r}_A} e^{i A_{\mathbf{r}_{A},\mathbf{r}_{A}+ \mathbf{b}_\mu} } \Phi_{\mathbf{r}_{A}+\mathbf{b}_\mu},
    \frac{1}{2} \Phi^{\dag}_{\mathbf{r}_A+\mathbf{b}_\mu} e^{-iA_{\mathbf{r}_{A},\mathbf{r}_{A}+ \mathbf{b}_\mu}}  \Phi_{\mathbf{r}_{A}},
    E_{\mathbf{r}_{A},\mathbf{r}_{A}+\mathbf{b}_\mu} }\nonumber \\ 
    &\mapsto   \qty\Big{ \frac{\gamma}{2}   \Phi^{\dag}_{\overline{C}_{6}(\mathbf{r}_A)} e^{i A_{\overline{C}_{6}(\mathbf{r}_{A}), \overline{C}_{6}(\mathbf{r}_{A}+\mathbf{b}_\mu)} } \Phi_{\overline{C}_{6}(\mathbf{r}_{A}+\mathbf{b}_\mu)} ,  \frac{\overline{\gamma}}{2} \Phi^{\dag}_{\overline{C}_{6}(\mathbf{r}_A+\mathbf{b}_\mu)} e^{-iA_{\overline{C}_{6}(\mathbf{r}_{A}),\overline{C}_{6}(\mathbf{r}_{A}+ \mathbf{b}_\mu)}} \Phi_{\overline{C}_{6}(\mathbf{r}_{A})} , E_{\overline{C}_{6}(\mathbf{r}_{A}),\overline{C}_{6}(\mathbf{r}_{A}+\mathbf{b}_\mu)}  } \\
    S :& \qty\Big{ \frac{1}{2} \Phi^{\dag}_{\mathbf{r}_A} e^{i A_{\mathbf{r}_{A},\mathbf{r}_{A}+ \mathbf{b}_\mu}}  \Phi_{\mathbf{r}_{A}+\mathbf{b}_\mu},
    \frac{1}{2} \Phi^{\dag}_{\mathbf{r}_A+\mathbf{b}_\mu} e^{-iA_{\mathbf{r}_{A},\mathbf{r}_{A}+ \mathbf{b}_\mu}}  \Phi_{\mathbf{r}_{A}},
    E_{\mathbf{r}_{A},\mathbf{r}_{A}+\mathbf{b}_\mu} }\nonumber \\ 
    &\mapsto   \qty\Big{  -\frac{\gamma}{2} \Phi^{\dag}_{S(\mathbf{r}_A+\mathbf{b}_\mu)} e^{-iA_{S(\mathbf{r}_{A}),S(\mathbf{r}_{A}+ \mathbf{b}_\mu)}}  \Phi_{S(\mathbf{r}_{A})},
    -\frac{\overline{\gamma}}{2}  \Phi^{\dag}_{S(\mathbf{r}_A)} e^{i A_{S(\mathbf{r}_{A}),S(\mathbf{r}_{A}+ \mathbf{b}_\mu)}}  \Phi_{S(\mathbf{r}_{A}+\mathbf{b}_\mu)} , -E_{S(\mathbf{r}_{A}),S(\mathbf{r}_{A}+\mathbf{b}_\mu)}}.
  \end{align}
\end{subequations}
\end{widetext}
With the vector notation introduced in Eq.~\eqref{eq: bosonic vector field}, we can rewrite these transformations as in Eq.~\eqref{eq: Transformation of the parton operators} accompanied by the gauge field transformations
\begin{subequations}  \label{eq: Transformation of the gauge field under SG}
    \begin{align}
    T_{i}: &
        A_{\mathbf{r}_{\alpha},\mathbf{r}_{\alpha}+\mathbf{b}_{\mu}} \mapsto A_{T_{i}(\mathbf{r}_{\alpha}), T_{i}(\mathbf{r}_{\alpha}+\mathbf{b}_{\mu})} \\
    \overline{C}_6: &
        A_{\mathbf{r}_{\alpha},\mathbf{r}_{\alpha}+\mathbf{b}_{\mu}} \mapsto A_{\overline{C}_6(\mathbf{r}_{\alpha}), \overline{C}_6(\mathbf{r}_{\alpha}+\mathbf{b}_{\mu})} + 2\pi/3 \\
    S: &
        A_{\mathbf{r}_{\alpha},\mathbf{r}_{\alpha}+\mathbf{b}_{\mu}} \mapsto -A_{S(\mathbf{r}_{\alpha}), S(\mathbf{r}_{\alpha}+\mathbf{b}_{\mu})} + 5\pi/3.
    \end{align}
\end{subequations}

\section{\label{appendix: Classification of symmetric $U(1)$ spin liquids} Classification of symmetric \texorpdfstring{$U(1)$}{U(1)} spin liquids}

\subsection{\label{appendix subsec: PSG solution - Generalities} Generalities}

To classify symmetry classes, one starts from all algebraic constraints of the form
\begin{equation} \label{eq: algebraic SG relations}
\mathcal{O}_{1} \circ \mathcal{O}_{2} \circ \cdots=1
\end{equation}
which translate directly to the gauge-enriched relations
\begin{equation} \label{eq: algebraic SG relations gauge-enriched}
\widetilde{\mathcal{O}}_{1} \circ \widetilde{\mathcal{O}}_{2} \circ \cdots=\left(G_{\mathcal{O}_{1}} \circ \mathcal{O}_{1}\right) \circ\left(G_{\mathcal{O}_{2}} \circ \mathcal{O}_{2}\right) \circ \cdots  = e^{i\psi} \in \text{IGG},
\end{equation}
with $\psi\in\left[0, 2\pi \right)$. We can use the following conjugation relation
\begin{widetext}
\begin{align}
\mathcal{O}_{i} \circ G_{\mathcal{O}_{j}} \circ \mathcal{O}_{i}^{-1}: \vec{\Psi}_{\mathbf{r}_{\alpha}} \mapsto \mqty( e^{i n_{\mathcal{O}_{i}} \phi_{\mathcal{O}_{j}}\left[\mathcal{O}_{i}^{-1}\left(\mathbf{r}_{\alpha}\right)\right]} & 0 \\ 0 & e^{-i n_{\mathcal{O}_{i}} \phi_{\mathcal{O}_{j}}\left[\mathcal{O}_{i}^{-1}\left(\mathbf{r}_{\alpha}\right)\right] } ) \vec{\Psi}_{\mathbf{r}_{\alpha}} = \left[\mathcal{G}_{\mathcal{O}_{j}}[\mathcal{O}_{i}^{-1}(\mathbf{r}_\alpha)]\right]^{n_{\mathcal{O}_{i}}} \vec{\Psi}_{\mathbf{r}_{\alpha}} ,
\end{align}
\end{widetext}
to map all these gauge-enriched constraints to phase relations of the form
\begin{equation}
\begin{aligned}\label{eq: phase equations for PSG}
&\phi_{\mathcal{O}_{1}}\left(\mathbf{r}_{\alpha}\right) +n_{\mathcal{O}_{1}}\phi_{\mathcal{O}_{2}}\left[\mathcal{O}_{1}^{-1}\left(\mathbf{r}_{\alpha}\right)\right] \\
&+ n_{\mathcal{O}_{1}} n_{\mathcal{O}_{2}} \phi_{\mathcal{O}_{3}}\left[\mathcal{O}_{2}^{-1}\circ\mathcal{O}_{1}^{-1}\left(\mathbf{r}_{\alpha}\right)\right]+\cdots= \psi\hspace{3mm} \text{mod }2\pi.
\end{aligned}
\end{equation}

The GMFT classes for a given IGG are then obtained by listing the gauge inequivalent solutions of all phase equations of the form~\eqref{eq: phase equations for PSG}. That is, it must be impossible to relate two distinct GMFT classes by a general gauge transformation $G$. Under such a gauge transformation, the phase factors are mapped to 
\begin{align}
    \phi_{\mathcal{O}}(\mathbf{r}_{\alpha}) &\to \phi_{\mathcal{O}}(\mathbf{r}_{\alpha}) + \phi_{G}(\mathbf{r}_{\alpha}) - n_{\mathcal{O}} \phi_{G}(\mathcal{O}^{-1}(\mathbf{r}_{\alpha}))
\end{align}
To identify inequivalent solutions, all gauge degrees of freedom must be fixed in the process of solving the algebraic equations. Considering spatially isotropic phase factors, there are two distinct gauge transformations for each sublattice ($\alpha\in\{A,B\}$ ) in every direction ($r_1$, $r_2$ and $r_3$)
\begin{align}
    G_{i,A}: \phi_{G_i,A}(\mathbf{r}_{\alpha}) &= \psi_{G_i,A} r_i \delta_{\alpha,A}, \\
    G_{i,B}: \phi_{G_i,B}(\mathbf{r}_{\alpha}) &= \psi_{G_i,B} r_i \delta_{\alpha,B}, 
\end{align}
one constant gauge transformations for every sublattice 
\begin{align}
    G_{A}^{\text{cst}}: \phi_{G_{A}^{\text{cst}}}(\mathbf{r}_{\alpha}) &= \psi_{A} \delta_{\alpha,A} \\
    G_{B}^{\text{cst}}: \phi_{G_{B}^{\text{cst}}}(\mathbf{r}_{\alpha}) &= \psi_{B} \delta_{\alpha,B},
\end{align}
where $\psi_{G_{i,\alpha}}$ and $\psi_{\alpha}$ are defined modulo $2\pi$. We are also free to add a site-independent phase factor to our five SG phases. Therefore, 8 local gauges and 6 phase factors must be fixed to get unambiguously inequivalent results. 

\subsection{\label{appendix subsec: PSG solution - Algebraic constraints} Algebraic constraints}

For the parent diamond lattice, the algebraic constraints are
\begin{subequations} \label{eq: SG generators constraints}
  \begin{align}
    T_{i} T_{i+1} T_{i}^{-1} T_{i+1}^{-1} &=1,  i=1,2,3 \\
    \bar{C}_{6}^{6} &=1, \\
    S^{2} T_{3}^{-1} &=1,  \\
    \bar{C}_{6} T_{i} \bar{C}_{6}^{-1} T_{i+1} &=1, i=1,2,3 \\
    S T_{i} S^{-1} T_{3}^{-1} T_{i} &=1,  i=1,2, \\
    S T_{3} S^{-1} T_{3}^{-1} &=1  \\
    \left(\bar{C}_{6} S\right)^{4} &=1 \\
    \left(\bar{C}_{6}^{3} S\right)^{2} &=1,  
  \end{align}
\end{subequations}
which correspond to the gauge-enriched operations
\begin{widetext}
\begin{subequations}
\begin{align}
\left(G_{T_{i}} T_{i}\right)\left(G_{T_{i+1}} T_{i+1}\right)\left(G_{T_{i}} T_{i}\right)^{-1}\left(G_{T_{i+1}} T_{i+1}\right)^{-1}&\in IGG, \\
\left(G_{\bar{C}_{6}} \bar{C}_{6}\right)^{6} &\in IGG, \\
\left(G_{S} S\right)^{2}\left(G_{T_{3}} T_{3}\right)^{-1} &\in IGG,\\
\left(G_{\bar{C}_{6}} \bar{C}_{6}\right)\left(G_{T_{i}} T_{i}\right)\left(G_{\bar{C}_{6}} \bar{C}_{6}\right)^{-1}\left(G_{T_{i+1}} T_{i+1}\right) &\in IGG, \\
\left(G_{S} S\right)\left(G_{T_{i}} T_{i}\right)\left(G_{S} S\right)^{-1}\left(G_{T_{3}} T_{3}\right)^{-1}\left(G_{T_{i}} T_{i}\right) &\in IGG, \\
\left(G_{S} S\right)\left(G_{T_{3}} T_{3}\right)\left(G_{S} S\right)^{-1}\left(G_{T_{3}} T_{3}\right)^{-1} &\in IGG, \\
\left[\left(G_{\bar{C}_{6}} \bar{C}_{6}\right)\left(G_{S} S\right)\right]^{4} &\in IGG,\\
\left[\left(G_{\bar{C}_{6}} \bar{C}_{6}\right)^{3}\left(G_{S} S\right)\right]^{2} &\in IGG .
\end{align}
\end{subequations}
In the case where $\text{IGG}=U(1)$, these constraints are explicitly 
\begin{subequations} 
\begin{empheq}[]{align}
\phi_{T_{i}}\left(\mathbf{r}_{\alpha}\right)+\phi_{T_{i+1}}\left[T_{i}^{-1}\left(\mathbf{r}_{\alpha}\right)\right]-\phi_{T_{i}}\left[T_{i+1}^{-1}\left(\mathbf{r}_{\alpha}\right)\right]-\phi_{T_{i+1}}\left(\mathbf{r}_{\alpha}\right) &=\psi_{T_i},\label{eq: psg classification T_i}\\
\phi_{\bar{C}_{6}}\left(\mathbf{r}_{\alpha}\right)+\phi_{\bar{C}_{6}}\left[\bar{C}_{6}^{-1}\left(\mathbf{r}_{\alpha}\right)\right]+\phi_{\bar{C}_{6}}\left[\bar{C}_{6}^{-2}\left(\mathbf{r}_{\alpha}\right)\right]+\phi_{\bar{C}_{6}}\left[\bar{C}_{6}^{-3}\left(\mathbf{r}_{\alpha}\right)\right]+\phi_{\bar{C}_{6}}\left[\bar{C}_{6}^{-4}\left(\mathbf{r}_{\alpha}\right)\right]+\phi_{\bar{C}_{6}}\left[\bar{C}_{6}^{-5}\left(\mathbf{r}_{\alpha}\right)\right] &=\psi_{\bar{C}_{6}} \label{eq: psg classification C} \\
\phi_{S}\left(\mathbf{r}_{\alpha}\right)-\phi_{S}\left[S^{-1}\left(\mathbf{r}_{\alpha}\right)\right]-\phi_{T_{3}}\left(\mathbf{r}_{\alpha}\right) &=\psi_{S} \label{eq: psg classification S}  \\
\phi_{\bar{C}_{6}}\left(\mathbf{r}_{\alpha}\right)+\phi_{T_{i}}\left[\bar{C}_{6}^{-1}\left(\mathbf{r}_{\alpha}\right)\right]-\phi_{\bar{C}_{6}}\left[T_{i+1}\left(\mathbf{r}_{\alpha}\right)\right]+\phi_{T_{i+1}}\left[T_{i+1}\left(\mathbf{r}_{\alpha}\right)\right] &=\psi_{\bar{C}_{6} T_{i}} \label{eq: psg classification CT_i}\\
\phi_{S}\left(\mathbf{r}_{\alpha}\right)-\phi_{T_{i}}\left[S^{-1}\left(\mathbf{r}_{\alpha}\right)\right]-\phi_{S}\left[T_{3}^{-1} T_{i}\left(\mathbf{r}_{\alpha}\right)\right]-\phi_{T_{3}}\left[T_{i}\left(\mathbf{r}_{\alpha}\right)\right]+\phi_{T_{i}}\left[T_{i}\left(\mathbf{r}_{\alpha}\right)\right]&=\psi_{S T_{i}} \label{eq: psg classification S T_i}  \\
\phi_{S}\left(\mathbf{r}_{\alpha}\right)-\phi_{T_{3}}\left[S^{-1}\left(\mathbf{r}_{\alpha}\right)\right]-\phi_{S}\left[T_{3}^{-1}\left(\mathbf{r}_{\alpha}\right)\right]-\phi_{T_{3}}\left(\mathbf{r}_{\alpha}\right) &=\psi_{S T_{3}} \label{eq: psg classification S T_3} \\
\phi_{\overline{C}_{6}}\left(\mathbf{r}_{\alpha}\right)+\phi_{S}\left[\bar{C}_{6}^{-1}\left(\mathbf{r}_{\alpha}\right)\right]-\phi_{\bar{C}_{6}}\left[\left(\bar{C}_{6} S\right)^{-1}\left(\mathbf{r}_{\alpha}\right)\right]-\phi_{S}\left[\left(\bar{C}_{6} S \bar{C}_{6}\right)^{-1}\left(\mathbf{r}_{\alpha}\right)\right]+\phi_{\bar{C}_{6}}\left[\left(\bar{C}_{6} S \bar{C}_{6} S\right)^{-1}\left(\mathbf{r}_{\alpha}\right)\right] & \nonumber\\
+\phi_{S}\left[\left(\bar{C}_{6} S \bar{C}_{6} S \bar{C}_{6}\right)^{-1}\left(\mathbf{r}_{\alpha}\right)\right]-\phi_{\bar{C}_{6}}\left[\left(\bar{C}_{6} S \bar{C}_{6} S \bar{C}_{6} S\right)^{-1}\left(\mathbf{r}_{\alpha}\right)\right]-\phi_{S}\left[\left(\bar{C}_{6} S \bar{C}_{6} S \bar{C}_{6} S \bar{C}_{6}\right)^{-1}\left(\mathbf{r}_{\alpha}\right)\right] &=\psi_{\bar{C}_{6} S} \label{eq: psg classification CS} \\
\phi_{\bar{C}_{6}}\left(\mathbf{r}_{\alpha}\right)+\phi_{\bar{C}_{6}}\left[\bar{C}_{6}^{-1}\left(\mathbf{r}_{\alpha}\right)\right]+\phi_{\bar{C}_{6}}\left[\bar{C}_{6}^{-2}\left(\mathbf{r}_{\alpha}\right)\right]+\phi_{S}\left[\bar{C}_{6}^{-3}\left(\mathbf{r}_{\alpha}\right)\right] -\phi_{\bar{C}_{6}}\left[\left(\bar{C}_{6}^{3} S\right)^{-1}\left(\mathbf{r}_{\alpha}\right)\right] \hspace{2.3cm} & \nonumber\\
-\phi_{\bar{C}_{6}}\left[\left(\bar{C}_{6}^{3} S \bar{C}_{6}\right)^{-1}\left(\mathbf{r}_{\alpha}\right)\right]-\phi_{\bar{C}_{6}}\left[\left(\bar{C}_{6}^{3} S \bar{C}_{6}^{2}\right)^{-1}\left(\mathbf{r}_{\alpha}\right)\right]-\phi_{S}\left[S\left(\mathbf{r}_{\alpha}\right)\right] &=\psi_{S \bar{C}_{6}}  \label{eq: psg classification SC}
\end{empheq}
\end{subequations}
\end{widetext}
where all $\psi\in\left[0,2\pi\right)$, $i=1,2,3$ for Eqs.~\eqref{eq: psg classification T_i} and~\eqref{eq: psg classification CT_i} and $i=1,2$ for Eq.~\eqref{eq: psg classification S T_i}. All phase equations are defined modulo $2\pi$. We will not indicate that subtlety explicitly for simplicity's sake.

\subsection{\label{appendix subsec: PSG solution - Solution of the PSG constraints} Solution of the constraints}

\subsubsection{\label{appendix subsec: PSG solution - Solution of the PSG constraints: inter-unit cell part} Inter-unit cell part}

Let us first consider the constraints coming from the commutativity of the translation operators given in Eq.~\eqref{eq: psg classification T_i}. Using our gauge freedom, we can set $\phi_{T_1}(r_1,r_2,r_3)_{\alpha}=\phi_{T_2}(0,r_2,r_3)_{\alpha}=\phi_{T1}(0,0,r_3)_{\alpha}=0$, which then leads to 
\begin{subequations} \label{eq U(1) classification: T1, T2, T3 first equation}
  \begin{align}
    \phi_{T_1}(\mathbf{r}_{\alpha}) &= 0 \\
    \phi_{T_2}(\mathbf{r}_{\alpha}) &= -\psi_{T_1} r_1\\
    \phi_{T_3}(\mathbf{r}_{\alpha}) &= \psi_{T_3} r_1 - \psi_{T_2} r_2.
  \end{align}
\end{subequations}
Plugging this into Eq.~\eqref{eq: psg classification CT_i}, we get
\begin{subequations}
\begin{align}
     \psi_{\overline{C}_6 T_1} =& \phi_{\overline{C}_6}(r_1,r_2,r_3)_{\alpha} -\phi_{\overline{C}_6}(r_1,r_2+1,r_3)_{\alpha} \nonumber \\
    &\quad\quad -r_1 \psi_{T_1} \\
    \psi_{\overline{C}_6 T_2} =&\phi_{\overline{C}_6}(r_1,r_2,r_3)_{\alpha} -\phi_{\overline{C}_6}(r_1,r_2,r_3+1)_{\alpha}  \nonumber \\
    &\quad \quad + \psi_{T_1} r_2 - \psi_{T_2} r_2 + \psi_{T_3} r_1 \\
    \psi_{\overline{C}_6 T_3}=&\phi_{\overline{C}_6}(r_1,r_2,r_3)_{\alpha} -\phi_{\overline{C}_6}(r_1+1,r_2,r_3)_{\alpha} \nonumber \\
    &\quad\quad + \psi_{T_2} r_3 - \psi_{T_3} r_2.
\end{align}
\end{subequations}
This yields $\psi_{T_1}=\psi_{T_2}=\psi_{T_3}$ and
\begin{align} \label{eq U(1) classification: first equation for phi C6}
    \phi_{\overline{C}_6}(\mathbf{r}_{\alpha}) =& \phi_{\overline{C}_6}(\mathbf{0}_{\alpha}) - r_2 \psi_{\overline{C}_6 T_1} -r_3 \psi_{\overline{C}_6 T_2} \nonumber \\
    &- r_1 \psi_{\overline{C}_6 T_{3}} - \psi_{T_1}(r_1 r_2 - r_1 r_3).
\end{align}
We can then replace the translation phase factors in the constraints~\eqref{eq: psg classification S T_i} and~\eqref{eq: psg classification S T_3} to find
\begin{subequations}
\begin{align}
    \psi_{S T_1} =&\phi_{S}(r_1,r_2,r_3)_{\alpha} -\phi_{S}(r_1+1,r_2,r_3-1)_{\alpha} \nonumber \\
    &\quad \quad+ (-1-r_1+r_2)\psi_{T_1}  \\
    \psi_{S T_2} =&\phi_{S}(r_1,r_2,r_3)_{\alpha} -\phi_{S}(r_1,r_2+1,r_3-1)_{\alpha} \nonumber \\
    &\quad \quad + (1-3 r_1+r_2) \psi_{T_1} \\
    \psi_{S T_3} =&\phi_{S}(r_1,r_2,r_3)_{\alpha} -\phi_{S}(r_1,r_2,r_3-1)_{\alpha}.
\end{align}
\end{subequations}
These equations impose $\psi_{T_1}=n_1 \pi$ with $n_1 \in\left\{0, 1 \right\}$ and 
\begin{align}\label{eq U(1) classification: first equation for phi S}
    \phi_{S}(\mathbf{r}_{\alpha}) =& \phi_{S}(\mathbf{0}_{\alpha}) -r_1 \psi_{ST_1} - r_2 \psi_{S T_2} \nonumber \\
    &+ \frac{1}{2} n_1 \pi \left( -r_1 + r_2 + 2 r_1 r_2 - r_1^2 + r_2^2  \right)\nonumber \\
    &+ (r_1 + r_2 + r_3)  \psi_{ST_3} .
\end{align}
Having the general form of the phase factors for the five space group generators, we can find all other constraints by replacing these in the remaining equations. The finite order of the rotoreflection $\overline{C}_6$ expressed in~\eqref{eq: psg classification C} leads to
\begin{subequations}
  \begin{align}
    3\phi_{\overline{C}_6}(\mathbf{0}_A) + 3\phi_{\overline{C}_6}(\mathbf{0}_B)  &= \psi_{\overline{C}_6}.
    \end{align}
\end{subequations}
Eq.~\eqref{eq: psg classification S} yields
\begin{subequations}
\begin{align}
    \psi_{S} =& -2 r_2 \psi_{ST_2} - 2 r_1 \psi_{S T_1} + ( r_1 + r_2) \psi_{S T_3} \nonumber \\
    &\quad + \phi_S(\mathbf{0}_A) - \phi_S(\mathbf{0}_B) \\
    \psi_S =& -2 r_2 \psi_{ST_2} - 2 r_1 \psi_{S T_1} + (1 + r_1 + r_2) \psi_{S T_3} \nonumber \\
    &\quad - \phi_S(\mathbf{0}_A) + \phi_S(\mathbf{0}_B) 
\end{align}
\end{subequations}
which leads to 
\begin{subequations}
  \begin{align}
    \psi_{ST_3} &= 2 \psi_{ST_1} \\
    \psi_{ST_1} &= \psi_{ST_2} \\
    \psi_{S} &= \psi_{ST_2} \\
    \phi_{S}(\mathbf{0}_A) - \phi_{S}(\mathbf{0}_B) &= \psi_{ST_2}.
  \end{align}
\end{subequations}
Eq.~\eqref{eq: psg classification CS} gives
\begin{subequations}
\begin{align}
     0 &= 2( \psi_{\overline{C}_6 T_1} - \psi_{\overline{C}_6 T_2} - \psi_{\overline{C}_6 T_3} - 2\psi_{ST_2} ) \\
    \psi_{\overline{C}_6 S} &=  \psi_{\overline{C}_6 T_1} - \psi_{\overline{C}_6 T_2} - \psi_{\overline{C}_6 T_3} - 2\psi_{ST_2}  
\end{align}
\end{subequations}
which is equivalent to
\begin{align}
     n_{\overline{C}_6 S} \pi &=  \psi_{\overline{C}_6 T_1} - \psi_{\overline{C}_6 T_2} - \psi_{\overline{C}_6 T_3} - 2\psi_{ST_2} 
\end{align}
with $n_{\overline{C}_6 S}\in \{0,1\}$. At last, Eq.~\eqref{eq: psg classification SC} gives
\begin{subequations}
  \begin{align}
    \psi_{S \overline{C}_{6}}  &= \psi_{\overline{C}_6 T_1} - \psi_{\overline{C}_6 T_2} - \psi_{\overline{C}_6 T_3} - 2 \psi_{ST_2} \\
    0 &= \psi_{\overline{C}_6 T_1} - \psi_{\overline{C}_6 T_2} - \psi_{\overline{C}_6 T_3} - 2 \psi_{ST_2} .
  \end{align}
\end{subequations}
Using the previous constraints, these imply
\begin{subequations}
  \begin{align}
    \psi_{\overline{C}_6 T_1} - \psi_{\overline{C}_6 T_2} - \psi_{\overline{C}_6 T_3} - 2 \psi_{ST_2} &= 0 \\
    n_{\overline{C}_6 S} &= 0 \\
    \psi_{S \overline{C}_6} &= 0.
  \end{align}
\end{subequations}

\subsubsection{\label{appendix subsec: PSG solution - Solution of the PSG constraints: intra-unit cell part} Gauge fixing and intra-unit cell part}

Now that all constraints coming from the space group have been determined, we need to fix all remaining gauge degrees of freedom and solve the intra-unit cell equations. Let us briefly summarize the results we have determined thus far. From the space group constraints we obtained the phase equations~\eqref{eq U(1) classification: T1, T2, T3 first equation},~\eqref{eq U(1) classification: first equation for phi C6} and~\eqref{eq U(1) classification: first equation for phi S}, and the constraints
\begin{subequations} \label{eq: intra-cell condition} 
\begin{align}
    3\phi_{\overline{C}_6}(\mathbf{0}_A) + 3\phi_{\overline{C}_6}(\mathbf{0}_B)  &= \psi_{\overline{C}_6} \label{eq: intra-cell condition Eq.1} \\
    \phi_{S}(\mathbf{0}_A)-  \phi_{S}(\mathbf{0}_B) &=\psi_{ST_{2}}  \label{eq: intra-cell condition Eq.2} \\
    \psi_{\overline{C}_6 T_1} - \psi_{\overline{C}_6 T_2} - \psi_{\overline{C}_6 T_3} - 2 \psi_{ST_2}&= 0. \label{eq: intra-cell condition Eq.3}
\end{align}
\end{subequations}
These constraints can be simplified by fixing some gauge degrees of freedom to remove redundant solutions. The phase associated with $T_1$, $T_2$ and $T_3$ appear an odd number of times in Eq.~\eqref{eq: psg classification CT_i}. Similarly, $T_3$ is also present an odd number of times in Eq.~\eqref{eq: psg classification S T_i}. Consequently, we can make use of our gauge freedom and IGG structure (i.e., $\phi_{\mathcal{O}}\to\phi_{\mathcal{O}}+\chi$, where $\chi \in\left[ 0,2\pi\right)$) for $\phi_{T_1}$, $\phi_{T_2}$ and $\phi_{T_3}$ to set $\psi_{\overline{C}_6 T_1}=\psi_{\overline{C}_6 T_2}=\psi_{S T_2}=0$. Such a gauge fixing also implies $\psi_{\overline{C}_6 T_3}=0$ from Eq.~\eqref{eq: intra-cell condition Eq.3}, and $\phi_{S}(\mathbf{0}_A) = \phi_{S}(\mathbf{0}_B)$ from Eq.~\eqref{eq: intra-cell condition Eq.2}. Next, we can use a constant sublattice-dependent gauge transformation of the form
\begin{equation} \label{eq: intra-cell condition sublattice gauge transformation}
    \phi(\mathbf{r}_{\alpha}) = \phi_{\alpha}, \hspace{5mm} \text{where }\alpha\in\{\text{A},\text{B}\}.
\end{equation}
As the phase factor transform according to $\phi_{\mathcal{O}}(\mathbf{r}_{\alpha})\to \phi_{\mathcal{O}}(\mathbf{r}_{\alpha}) + \phi(\mathbf{r}_{\alpha}) - n_{\mathcal{O}}\phi\left[ \mathcal{O}^{-1}(\mathbf{r}_{\alpha}) \right]$ for a general gauge transformation, our initial gauge fixing for $\phi_{T_1}$, $\phi_{T_{2}}$ and $\phi_{T_{3}}$ are unaffected by the gauge transformation of Eq.~\eqref{eq: intra-cell condition sublattice gauge transformation} while $\phi_{\overline{C}_6}$ and $\phi_{S}$ are mapped to
\begin{subequations}
  \begin{align}
    \phi_{\overline{C}_6}(\mathbf{0}_{\alpha}) &\to \eta_{\alpha}  (\phi_{A} - \phi_{B}) +  \phi_{\overline{C}_6}(\mathbf{0}_{\alpha}) \\
    \phi_{S}(\mathbf{0}_{\alpha}) &\to  (\phi_{A} + \phi_{B}) +  \phi_{S}(\mathbf{0}_{\alpha}) 
  \end{align}
\end{subequations}
We can then choose $\phi_{\alpha}$ and make use of our IGG freedom for $\phi_{\overline{C}_6}$ and $\phi_{S}$ to fix 
\begin{equation}
    \phi_{\overline{C}_{6}} (\mathbf{0}_B) = \phi_{\overline{C}_6}(\mathbf{0}_A) = \phi_{S}(\mathbf{0}_A)  = 0
\end{equation}
This implies that $\psi_{\overline{C}_6}=0$ from Eq.~\eqref{eq: intra-cell condition Eq.1}.  

We conclude that there are only two GMFT classes given by the phase factors summarized in Eq.~\eqref{eq: U(1) PSG classification with TRS and inversion}.

\section{\label{appendix: Classification of chiral $U(1)$ spin liquids} Classification of chiral \texorpdfstring{$U(1)$}{U(1)} spin liquids}

\subsection{\label{appendix subsec: chiral PSG solution - even subgroup} Even subgroup}

We are interested in finding the even subgroup $\chi_e \subseteq\text{SG}$ of transformations. To do so, a parity is associated with every transformation. This parity indicates if the GMFT \emph{Ansatz} respects that symmetry directly (i.e., $\epsilon_{\mathcal{O}}=1$ for all $\mathcal{O}\in\chi_{e}$) or modulo a time-reversal operation (i.e., $\epsilon_{\mathcal{O}}=-1$ for all $\mathcal{O}\in\chi_{o}$). It is first trivial to notice that, since $\epsilon^2=1$ for $\epsilon=\pm 1$, all SG generator squared are elements of the even subgroup $\{T_{1}^2$, $T_{2}^2$, $T_{3}^2$, $S^2$, $\overline{C}_{6}^2=C_{3}^{-1}\}$ $\in\chi_{e}$. Next, all SG algebraic constraints expressed in Eq.~\eqref{eq: SG generators constraints} can be translated into the following equations for the parity of the SG generators 
\begin{subequations} \label{eq: SG generators parity constraints}
  \begin{align}
    \epsilon_{T_{i}} \epsilon_{T_{i+1}} \epsilon_{T_{i}} \epsilon_{T_{i+1}} &=1,  i=1,2,3 \\
    \epsilon_{\bar{C}_{6}}^{6} &=1, \\
    \epsilon_{S}^{2} \epsilon_{T_{3}} &=1,  \label{eq: SG generator parity NT 1} \\
    \epsilon_{\bar{C}_{6}} \epsilon_{T_{i}} \epsilon_{\bar{C}_{6}} \epsilon_{T_{i+1}} &=1, i=1,2,3  \label{eq: SG generator parity NT 2} \\
    \epsilon_{S} \epsilon_{T_{i}} \epsilon_{S} \epsilon_{T_{3}} \epsilon_{T_{i}} &=1,  i=1,2, \label{eq: SG generator parity NT 3}  \\
    \epsilon_{S} \epsilon_{T_{3}} \epsilon_{S} \epsilon_{T_{3}} &=1  \\
    \left(\epsilon_{\bar{C}_{6}} \epsilon_{S} \right)^{4} &=1 \\
    \left(\epsilon_{\bar{C}_{6}}^{3} \epsilon_{S}\right)^{2} &=1.
  \end{align}
\end{subequations}
Most of these equations are trivial. However, Eqs.~\eqref{eq: SG generator parity NT 1},~\eqref{eq: SG generator parity NT 2}, and~\eqref{eq: SG generator parity NT 3} imply $\epsilon_{T_{1}}=\epsilon_{T_{2}}=\epsilon_{T_{3}}=1$ while $\overline{C}_6$ and $S$ remain of undetermined parity. From this point, new generators of $\chi_{e}$ can be found by using the fact that
\begin{equation}
    \mathcal{O}_{o}^{-1} \mathcal{O}_{e} \mathcal{O}_{o} \in \chi_{e}
\end{equation}
for any $\mathcal{O}_{o}$ and $\mathcal{O}_{e}\in \chi_{e}$. We can then proceed iteratively for any $\mathcal{O}_{e}$ and $\mathcal{O}_{o}$ until no new generators of $\chi_{e}$ are produced. In this case, the only new generator of $\chi_{e}$ that can be found this way is 
\begin{align} \label{eq: definition C3 prime}
    C_{3}^\prime = S^{-1} C_{3} S = S^{-1} (\overline{C}_{6}^{4}) S.
\end{align}
In summary, $\{T_{1}$, $T_{2}$, $T_{3}$, $C_{3}$, $C_{3}^\prime\}\in\chi_{e}$ are the generators for the even subgroup of spatial transformations.

\subsection{\label{appendix subsec: chiral PSG solution - Algebraic constraints} Algebraic constraints}

The algebraic constraints on the even subgroup generators determined in Appendix~\ref{appendix subsec: chiral PSG solution - even subgroup} are
\begin{subequations} \label{eq: chiral SG generators constraints}
  \begin{align}
    T_{i} T_{i+1} T_{i}^{-1} T_{i+1}^{-1} &=1, i=1,2,3 \\
    C_{3}^{3} &=1 \\
    C_{3}^{\prime 3} &=1 \\
    \left(C_{3} C_{3}^{\prime}\right)^{2} &=1 \\
    C_{3} T_{i} C_{3}^{-1} T_{i+1}^{-1} &=1 , i=1,2,3 \\
    C_{3}^{\prime} T_{1}\left(C_{3}^{\prime}\right)^{-1} T_{1} T_{2}^{-1} &=1 \\
    C_{3}^{\prime} T_{2}\left(C_{3}^{\prime}\right)^{-1} T_{1} &=1 \\
    C_{3}^{\prime} T_{3}\left(C_{3}^{\prime}\right)^{-1} T_{1} T_{3}^{-1} &=1.
  \end{align}
\end{subequations}
These correspond to the following gauge-enriched operations
\begin{widetext}
\begin{subequations}
\begin{align}
\left(G_{T_{i}} T_{i}\right)\left(G_{T_{i+1}} T_{i+1}\right)\left(G_{T_{i}} T_{i}\right)^{-1}\left(G_{T_{i+1}} T_{i+1}\right)^{-1}  &\in IGG \\
\left(G_{C_{3}} C_{3}\right)^{3} &\in IGG \\
\left(G_{C_{3}^{\prime}} C_{3}^{\prime}\right)^{3} &\in IGG \\
\left(G_{C_{3}} C_{3}\right)\left(G_{C_{3}^{\prime}} C_{3}^{\prime}\right)\left(G_{C_{3}} C_{3}\right)\left(G_{C_{3}^{\prime}} C_{3}^{\prime}\right) &\in IGG \\
\left(G_{C_{3}} C_{3}\right)\left(G_{T_{i}} T_{i}\right)\left(G_{C_{3}} C_{3}\right)^{-1}\left(G_{T_{i+1}} T_{i+1}\right)^{-1} &\in IGG \\
\left(G_{C_{3}^{\prime}} C_{3}^{\prime}\right)\left(G_{T_{1}} T_{1}\right)\left(G_{C_{3}^{\prime}} C_{3}^{\prime}\right)^{-1} \left(G_{T_{1}} T_{1}\right) \left(G_{T_{2}} T_{2}\right)^{-1} &\in IGG \\
\left(G_{C_{3}^{\prime}} C_{3}^{\prime}\right)\left(G_{T_{2}} T_{2}\right)\left(G_{C_{3}^{\prime}} C_{3}^{\prime}\right)^{-1}\left(G_{T_{1}} T_{1}\right) &\in IGG \\
\left(G_{C_{3}^{\prime}} C_{3}^{\prime}\right)\left(G_{T_{3}} T_{3}\right)\left(G_{C_{3}^{\prime}} C_{3}^{\prime}\right)^{-1} \left(G_{T_{1}} T_{1}\right) \left(G_{T_{3}} T_{3}\right)^{-1} &\in IGG.
\end{align}
\end{subequations}
When $\text{IGG}=U(1)$, these constraints lead to the following phase equations
\begin{subequations} 
\begin{align}
\phi_{T_{i}}\left[\mathbf{r}_{\alpha}\right]+\phi_{T_{i+1}}\left[T_{i}^{-1}\left(\mathbf{r}_{\alpha}\right)\right]-\phi_{T_{i}}\left[T_{i+1}^{-1}\left(\mathbf{r}_{\alpha}\right)\right]-\phi_{T_{i+1}}\left[\mathbf{r}_{\alpha}\right] &= \psi_{T_{i}} \label{eq: chiral psg classification Ti} \\
\phi_{C_{3}}\left[\mathbf{r}_{\alpha}\right]+\phi_{C_{3}}\left[C_{3}^{2}\left(\mathbf{r}_{\alpha}\right)\right]+\phi_{C_{3}}\left[C_{3}\left(\mathbf{r}_{\alpha}\right)\right] &= \psi_{C_{3}} \label{eq: chiral psg classification C3} \\
\phi_{C_{3}^{\prime}}\left[\mathbf{r}_{\alpha}\right]+\phi_{C_{3}^{\prime}}\left[\left(C_{3}^{\prime}\right)^{2}\left(\mathbf{r}_{\alpha}\right)\right]+\phi_{C_{3}^{\prime}}\left[\left(C_{3}^{\prime}\right)\left(\mathbf{r}_{\alpha}\right)\right] &= \psi_{C_{3}^{\prime}}, \label{eq: chiral psg classification C3p} \\
\phi_{C_{3}}\left[\mathbf{r}_{\alpha}\right]+\phi_{C_{3}^{\prime}}\left[\left(C_{3}\right)^{-1}\left(\mathbf{r}_{\alpha}\right)\right]+\phi_{C_{3}}\left[\left(C_{3} C_{3}^{\prime}\right)\left(\mathbf{r}_{\alpha}\right)\right]+\phi_{C_{3}^{\prime}}\left[C_{3}^{\prime}\left(\mathbf{r}_{\alpha}\right)\right] &= \psi_{C_{3} C_{3}^{\prime}}, \label{eq: chiral psg classification C3C3p} \\
\phi_{C_{3}}\left[\mathbf{r}_{\alpha}\right]+\phi_{T_{i}}\left[C_{3}^{-1}\left(\mathbf{r}_{\alpha}\right)\right]-\phi_{C_{3}}\left[T_{i+1}^{-1}\left(\mathbf{r}_{\alpha}\right)\right]-\phi_{T_{i+1}}\left[\mathbf{r}_{\alpha}\right] &= \psi_{C_{3} T_{i}} \label{eq: chiral psg classification C3Ti} \\
\phi_{C_{3}^{\prime}}\left[\mathbf{r}_{\alpha}\right]+\phi_{T_{1}}\left[\left(C_{3}^{\prime}\right)^{-1}\left(\mathbf{r}_{\alpha}\right)\right]-\phi_{C_{3}^{\prime}}\left[T_{1} T_{2}^{-1}\left(\mathbf{r}_{\alpha}\right)\right]+\phi_{T_{1}}\left[T_{1}T_{2}^{-1}\left(\mathbf{r}_{\alpha}\right)\right]-\phi_{T_{2}}\left[\mathbf{r}_{\alpha}\right] &= \psi_{C_{3}^{\prime} T_{1}} \label{eq: chiral psg classification C3pT1}\\
\phi_{C_{3}^{\prime}}\left[\mathbf{r}_{\alpha}\right]+\phi_{T_{2}}\left[\left(C_{3}^{\prime}\right)^{-1}\left(\mathbf{r}_{\alpha}\right)\right]-\phi_{C_{3}^{\prime}}\left[T_{1}\left(\mathbf{r}_{\alpha}\right)\right]+\phi_{T_{1}}\left[T_{1}\left(\mathbf{r}_{\alpha}\right)\right] &= \psi_{C_{3}^{\prime} T_{2}}  \label{eq: chiral psg classification C3pT2} \\
\phi_{C_{3}^{\prime}}\left[\mathbf{r}_{\alpha}\right]+\phi_{T_{3}}\left[\left(C_{3}^{\prime}\right)^{-1}\left(\mathbf{r}_{\alpha}\right)\right]-\phi_{C_{3}^{\prime}}\left[T_{1} T_{3}^{-1}\left(\mathbf{r}_{\alpha}\right)\right]+\phi_{T_{1}}\left[T_{1}T_{3}^{-1}\left(\mathbf{r}_{\alpha}\right)\right]-\phi_{T_{3}}\left[\mathbf{r}_{\alpha}\right] &= \psi_{C_{3}^{\prime} T_{3}}  \label{eq: chiral psg classification C3pT3}
\end{align}
\end{subequations}
\end{widetext}
where all $\psi\in\left[0,2\pi\right)$ and $i=1,2,3$ for Eqs.~\eqref{eq: chiral psg classification Ti} and~\eqref{eq: chiral psg classification C3Ti}.

\subsection{\label{appendix subsec: Chiral PSG solution - Solution of the PSG constraints} Solution of the constraints}

\subsubsection{\label{appendix subsec: Chiral PSG solution - Solution of the PSG constraints: inter-unit cell part} Inter-unit cell part}

Proceeding in a similar way to the fully symmetric classification, we can use our gauge freedom to set $\phi_{T_1}(r_1,r_2,r_3)_{\alpha}=\phi_{T_2}(0,r_2,r_3)_{\alpha}=\phi_{T1}(0,0,r_3)_{\alpha}=0$. This leads to 
\begin{subequations} \label{eq: chiral U(1) classification: T1, T2, T3 first equation}
  \begin{align}
    \phi_{T_1}(\mathbf{r}_{\alpha}) &= 0 \\
    \phi_{T_2}(\mathbf{r}_{\alpha}) &= -\psi_{T_1} r_1\\
    \phi_{T_3}(\mathbf{r}_{\alpha}) &= \psi_{T_3} r_1 - \psi_{T_2} r_2.
  \end{align}
\end{subequations}
From Eq.~\eqref{eq: chiral psg classification C3Ti}, we get 
\begin{subequations}
\begin{align}
    \psi_{C_3 T_1} =&  \phi_{C_{3}}(r_1, r_2, r_3)_{\alpha} - \phi_{C_{3}}(r_1, r_2-1, r_3)_{\alpha}\nonumber \\
    &\quad + r_1 \psi_{T_1} \\
    \psi_{C_3 T_2} =& \phi_{C_{3}}(r_1, r_2, r_3)_{\alpha} - \phi_{C_{3}}(r_1, r_2, r_3-1)_{\alpha} \nonumber \\
    &\quad - r_2 \psi_{T_1} + r_2 \psi_{T_2} - r_1 \psi_{T_3}  \\
    \psi_{C_3 T_3} =& \phi_{C_{3}}(r_1, r_2, r_3)_{\alpha} - \phi_{C_{3}}(r_1-1, r_2, r_3)_{\alpha} \nonumber \\
    &\quad - r_3 \psi_{T_2} + r_2 \psi_{T_3}
\end{align}
\end{subequations}
which enforces $\psi_{T_1} = \psi_{T_2} = \psi_{T_3}$ and 
\begin{align} \label{eq: chiral U(1) classification: C3 first equation}
    \phi_{C_{3}}(\mathbf{r}_\alpha) =& \phi_{C_{3}}(\mathbf{0}_\alpha) + r_2 \psi_{C_3 T_1} + r_3 \psi_{C_3 T_2} \nonumber \\
    &\quad + r_1 \psi_{C_3 T_3} - \psi_{T_1} (r_1 r_2 - r_1 r_3).
\end{align}
Replacing the translation phase factors in Eqs. \eqref{eq: chiral psg classification C3pT1}-\eqref{eq: chiral psg classification C3pT3} gives
\begin{subequations}
\begin{align}
    \psi_{C_3^\prime T_1} =& \phi_{C_{3}^\prime}(r_1, r_2, r_3)_{\alpha} - \phi_{C_{3}^\prime}(r_1+1, r_2-1, r_3)_{\alpha} \nonumber \\
    &\quad + r_1 \psi_{T_1}  \\
    \psi_{C_3^\prime T_2} =& \phi_{C_{3}^\prime}(r_1, r_2, r_3)_{\alpha} - \phi_{C_{3}^\prime}(r_1+1, r_2, r_3)_{\alpha} \nonumber \\
    &\quad- r_2 \psi_{T_1} \\
    \psi_{C_3^\prime T_3} =& \phi_{C_{3}^\prime}(r_1, r_2, r_3)_{\alpha} - \phi_{C_{3}}(r_1+1, r_2, r_3-1)_{\alpha} \nonumber \\
    &\quad + (\delta_{\alpha,0} + 3 r_2 + r_3).
\end{align}
\end{subequations}
Solving these equations, we get that $4 \psi_{T_1} = 0$ which implies $\psi_{T_1}= \frac{ n_{1/2} \pi}{2}$ for $n_{1/2}\in\{0,1,2,3\}$ and 
\begin{align} \label{eq: chiral U(1) classification:  C3p first equation}
    \psi_{C_{3}^\prime} (\mathbf{r}_\alpha) =& \frac{-n_{1}\pi}{4} \left( r_2 (2r_1+r_2 -1) \right) + r_3 (1+2\delta_{\alpha,0}+r_3) \nonumber \\
    &+ r_2 \psi_{C_{3}^\prime T_1} - (r_1 + r_2 + r_3) \psi_{C_3^\prime T_2} + r_3 \psi_{C_3^\prime T_3}.
\end{align}
The finite order of the $C_3$ operation expressed in Eq.~\eqref{eq: chiral psg classification C3} gives the equation
\begin{align}
    \psi_{C_3} =&  (r_{1} + r_{2} + r_{3}) (\psi_{C_{3} T_1} + \psi_{C_{3} T_2} + \psi_{C_{3} T_3}) \nonumber \\
    &\quad + 3 \phi_{C_3}(\mathbf{0}_\alpha) 
\end{align}
which leads to the constraints
\begin{subequations}
\begin{align}
    \psi_{C_{3} T_{1}} + \psi_{C_{3} T_{2}} + \psi_{C_{3} T_{3}} &= 0\\
    3 \phi_{C_{3}}(\mathbf{0}_{\alpha}) &= \psi_{C_3} \text{ for } \alpha\in\{\text{A},\text{B}\}.
\end{align}
\end{subequations}
Similarly, Eq.~\eqref{eq: chiral psg classification C3p} associated with the finite order of $C_{3}^{\prime}$ imposes
\begin{align}
    \psi_{C_3} =& -r_3 (\psi_{C_{3}^\prime T_1} + \psi_{C_{3^\prime} T_2} -3 \psi_{C_{3}^\prime T_3}) + 3 \phi_{C_3^\prime}(\mathbf{0}_\alpha)   \nonumber \\
    &\quad + \left( -\psi_{C_3^\prime T_1} + 2 \psi_{C_{3}^\prime T_2} - \frac{n_{1/2} \pi}{2} \right) \delta_{\alpha,0}.
\end{align}
After solving the finite difference equation, one finds
\begin{subequations}
\begin{align}
    0 =& \psi_{C_{3}^\prime T_{1}} + \psi_{C_{3}^\prime T_{2}} - 3 \psi_{C_{3}^\prime T_{3}} \\
    \psi_{C_3^\prime} =& - \psi_{C_3^\prime T_1} + 2 \psi_{C_3^\prime T_2} - \frac{n_{1/2}\pi}{2} + 3 \phi_{C_3^\prime} (\mathbf{0}_{A})  \\
    \psi_{C_3^\prime} =& 3 \phi_{C_{3}^\prime} (\mathbf{0}_{B}).
\end{align}
\end{subequations}
Replacing all the space group generator phase factors in Eq.~\eqref{eq: chiral psg classification C3C3p} results in
\begin{align}
    \psi_{C_3 C_3^\prime} =& (r_1 + r_3) \left(\psi_{C_{3}^\prime T_{1}} - \psi_{C_{3}^\prime T_{2}} + \psi_{C_{3}^\prime T_{3}} - \psi_{C_3 T_1} \right. \nonumber \\
    &\left. + \psi_{C_3 T_2} + \psi_{C_3 T_3} \right) + (\psi_{C_3^\prime T_2} - \psi_{C_3 T_1} )\delta_{\alpha,0} \nonumber \\
    &+ 2 ( \phi_{C_{3}}(\mathbf{0}_\alpha) + \phi_{C_{3}^\prime}(\mathbf{0}_\alpha)  ) 
\end{align}
which is equivalent to the constraints
\begin{subequations}
\begin{align}
    0 =& \psi_{C_{3}^\prime T_{1}} - \psi_{C_{3}^\prime T_{2}} + \psi_{C_{3}^\prime T_{3}} \nonumber \\
    &\quad - \psi_{C_3 T_1} + \psi_{C_3 T_2} + \psi_{C_3 T_3}  \\
    \psi_{C_3 C_3^\prime} =& (\psi_{C_3^\prime T_2} - \psi_{C_3 T_1} ) \nonumber \\
    &\quad + 2 ( \phi_{C_{3}}(\mathbf{0}_A) + \phi_{C_{3}^\prime}(\mathbf{0}_A)  ) \\
     \psi_{C_3 C_3^\prime} =& 2 ( \phi_{C_{3}}(\mathbf{0}_B) + \phi_{C_{3}^\prime}(\mathbf{0}_B)  ).
\end{align}
\end{subequations}

\subsubsection{\label{appendix subsec: Chiral PSG solution - Solution of the PSG constraints: intra-unit cell part} Gauge fixing and intra-unit cell part}

In summary,  we have the phase equations 
~\eqref{eq: chiral U(1) classification:  C3 first equation},~\eqref{eq: chiral U(1) classification: T1, T2, T3 first equation},~\eqref{eq: chiral U(1) classification:  C3 first equation} and~\eqref{eq: chiral U(1) classification:  C3p first equation}, with the following constraints
\begin{subequations}
\begin{align}
    0 =& \psi_{C_{3} T_{1}} + \psi_{C_{3} T_{2}} + \psi_{C_{3} T_{3}}  \label{eq: chiral classification, constraint 1} \\
    \psi_{C_3} =& 3 \phi_{C_{3}}(\mathbf{0}_{A})  \label{eq: chiral classification, constraint 2} \\
    \psi_{C_3} =& 3 \phi_{C_{3}}(\mathbf{0}_{B})  \label{eq: chiral classification, constraint 3} \\
    0 =& \psi_{C_{3}^\prime T_{1}} + \psi_{C_{3}^\prime T_{2}} - 3 \psi_{C_{3}^\prime T_{3}} \label{eq: chiral classification, constraint 4}  \\
    \psi_{C_3^\prime} =& - \psi_{C_3^\prime T_1} + 2 \psi_{C_3^\prime T_2} 
    - \frac{n_{1/2}\pi}{2} + 3 \phi_{C_3^\prime} (\mathbf{0}_{A}) \label{eq: chiral classification, constraint 5}  \\
    \psi_{C_3^\prime} =& 3 \phi_{C_{3}^\prime} (\mathbf{0}_{B})  \label{eq: chiral classification, constraint 6} \\
    0 =& \psi_{C_{3}^\prime T_{1}} - \psi_{C_{3}^\prime T_{2}} + \psi_{C_{3}^\prime T_{3}} \nonumber \\
    &\quad - \psi_{C_3 T_1} + \psi_{C_3 T_2} + \psi_{C_3 T_3}  \label{eq: chiral classification, constraint 7} \\
    \psi_{C_3 C_3^\prime} =& (\psi_{C_3^\prime T_2} - \psi_{C_3 T_1} ) + 2 ( \phi_{C_{3}}(\mathbf{0}_A) + \phi_{C_{3}^\prime}(\mathbf{0}_A)  ) \label{eq: chiral classification, constraint 8}  \\
     \psi_{C_3 C_3^\prime} =& 2 ( \phi_{C_{3}}(\mathbf{0}_B) + \phi_{C_{3}^\prime}(\mathbf{0}_B)  )  \label{eq: chiral classification, constraint 9}.
\end{align}
\end{subequations}
First, from Eqs.~\eqref{eq: chiral classification, constraint 1} and~\eqref{eq: chiral classification, constraint 4} 
\begin{subequations}
\begin{align}
    \psi_{C_3 T_1} &= -\psi_{C_3 T_2} - \psi_{C_3 T_3} \\
    \psi_{C_3^\prime T_1} &= - \psi_{C_3^\prime T_2} + 3 \psi_{C_3^\prime T_3}.
\end{align}
\end{subequations}
Next, from Eqs.~\eqref{eq: chiral classification, constraint 2} and~\eqref{eq: chiral classification, constraint 3} 
\begin{subequations}
\begin{align}
    \phi_{C_3}(\mathbf{0}_A) &= \frac{\psi_{C_3}}{3} \\
    \phi_{C_3}(\mathbf{0}_B) &= \frac{\psi_{C_3}}{3}.
\end{align}
\end{subequations}
We can use our gauge freedom for $\phi_{T_1}$, $\phi_{T_2}$ and $\phi_{T_3}$ to fix $\psi_{C_3 T_2} = \psi_{C_3 T_3} = \psi_{C_3^\prime T_2}=0$ since the phase factors for $T_{1}$, $T_2$ and $T_3$ appear an odd number of times in Eq.~\eqref{eq: chiral psg classification C3Ti}. Next, with Eq.~\eqref{eq: chiral classification, constraint 7} we get
\begin{align}
    \psi_{C_3' T_3} &= \frac{n_{C_{3}^\prime T_3} \pi}{2},
\end{align}
where $n_{C_{3}^\prime T_3} \in \{0,1,2,3\}$. We can then use our IGG degree of freedom to fix $\psi_{C_3}=0$ because the phase factor for $C_3$ appears an odd number of times in Eq.~\eqref{eq: chiral psg classification C3}. The sublattice-dependent constant gauge degree of freedom $\phi_{G^{\text{cst}}_{\alpha}}(\mathbf{r}_\alpha) = \psi_{\beta} \delta_{\alpha \beta}$ can also be fixed. Under such a gauge transformation, the phase factors of the symmetry generators transform as 
\begin{subequations}
\begin{align}
    \phi_{T_i}(\mathbf{r}_\alpha) &\to \phi_{T_i}(\mathbf{r}_\alpha)  \\
    \phi_{C_3}(\mathbf{r}_\alpha) &\to \phi_{C_3}(\mathbf{r}_\alpha) \\
    \phi_{C_3^\prime}(\mathbf{r}_\alpha) &\to \phi_{C_3^\prime}(\mathbf{r}_\alpha) + 2 (\psi_{A} \delta_{\alpha,A} + \psi_{B} \delta_{\alpha,B} ).
\end{align}
\end{subequations}
Therefore, we may use it to fix $\phi_{C_{3}^\prime}(\mathbf{0}_A) = \phi_{C_{3}^\prime}(\mathbf{0}_B) = 0$. Eqs.~\eqref{eq: chiral classification, constraint 6},~\eqref{eq: chiral classification, constraint 8} and~\eqref{eq: chiral classification, constraint 9} now directly imply $\psi_{C_{3}^\prime}=\psi_{C_{3} C_{3}^\prime}=0$. Finally, Eq.~\eqref{eq: chiral classification, constraint 5} yields $n_{C_{3}^\prime T_{3}} = n_{1/2}$. 

In conclusion, we find four GMFT classes characterized by the phase factors of Eq.~\eqref{eq: U(1) chiral PSG classification}.

\section{\label{appendix :  Relation between MF parameters on different bonds in the unit cell} Relation between gauge field on different bonds}

As explained in Sec.~\ref{subsec: Gauge fields configurations}, 
we need to pick a specific representative bond and then map it to all other bonds of the lattice to find the gauge field configuration for all \emph{Ans\"atze}. We take the bond $(\mathbf{0}_A \to \mathbf{0}_B) $ to be the representative bond of reference and set the corresponding gauge field to an arbitrary value $\overline{A}$. Then we can find a mapping from it to the three other bonds of the primitive diamond unit cell: $(\mathbf{0}_A \to (1,0,0)_B)$, $(\mathbf{0}_A \to (0,1,0)_B)$, and $(\mathbf{0}_A \to (0,0,1)_B)$. All other lattice bonds can be obtained by iterative application of the translation operators from these four primitive bonds. 

\subsection{\label{appendix subsec: Relation between MF parameters on different bonds in the unit cell -> Symmetric case} Symmetric classification}

For the fully symmetric case, the four bonds of the diamond lattice primitive unit cell are related by the following composition of the symmetry generators
\begin{subequations}
\begin{align}
E:& (\mathbf{0}_A \to \mathbf{0}_B) \mapsto (\mathbf{0}_A \to \mathbf{0}_B) \\
\overline{C}_6^{4}\circ S\circ\overline{C}_6^{3}:& (\mathbf{0}_A \to \mathbf{0}_B) \mapsto (\mathbf{0}_A \to (1,0,0)_B)  \\
\overline{C}_6^{2} \circ S \circ\overline{C}_6^{3} :& (\mathbf{0}_A \to \mathbf{0}_B) \mapsto (\mathbf{0}_A \to (0,1,0)_B)  \\
 S \circ\overline{C}_6^{3}:& (\mathbf{0}_A \to \mathbf{0}_B) \mapsto (\mathbf{0}_A \to (0,0,1)_B).
\end{align}
\end{subequations}
From these transformations, the mapping of the gauge field between different bonds expressed in Eq.~\eqref{eq: mapping of the gauge fields under SG operations}, and the phases in Eq.~\eqref{eq: U(1) PSG classification with TRS and inversion}, we get the following relations for the gauge field on all bonds of the diamond lattice
\begin{subequations}
\begin{align}
\overline{A}_{(0,0,0)_{A},(0,0,0)_{B}} &= \overline{A}  \\
\overline{A}_{(r_{1},r_{2},r_{3})_{A},(r_{1},r_{2},r_{3})_{B}} &= \overline{A}  \\
\overline{A}_{(r_{1},r_{2},r_{3})_{A},(r_{1}+1,r_{2},r_{3})_{B}} &= -\overline{A} + n_{1} \pi (r_{2} + r_{3}) \\
\overline{A}_{(r_{1},r_{2},r_{3})_{A},(r_{1},r_{2}+1,r_{3})_{B}} &= -\overline{A} + n_{1} \pi r_{3} \\
\overline{A}_{(r_{1},r_{2},r_{3})_{A},(r_{1},r_{2},r_{3}+1)_{B}} &= -\overline{A}.
\end{align}
\end{subequations}

\subsection{\label{appendix subsubsec: Relation between MF parameters on different bonds in the unit cell -> Chiral case} Chiral classification}

The quotient group associated with the even subgroup (i.e., $\chi_e/\mathbf{T}_{0}$ where $\mathbf{T}_0$ is the abelian normal subgroup of translations generated by $T_1$, $T_2$ and $T_3$) has two generators $C_3$ and $T_1\circ C_{3}^\prime$. The following operations relate the bonds of the primitive diamond unit cell
\begin{subequations}
\begin{align}
E:& (\mathbf{0}_A \to \mathbf{0}_B) \mapsto (\mathbf{0}_A \to \mathbf{0}_B) \\
T_1\circ C_{3}^\prime  :& (\mathbf{0}_A \to \mathbf{0}_B) \mapsto (\mathbf{0}_A \to (1,0,0)_B)  \\
(T_1\circ C_{3}^\prime)^2  :& (\mathbf{0}_A \to \mathbf{0}_B) \mapsto (\mathbf{0}_A \to (0,1,0)_B)  \\
C_3 \circ (T_1\circ C_{3}^\prime)^2 :& (\mathbf{0}_A \to \mathbf{0}_B) \mapsto (\mathbf{0}_A \to (0,0,1)_B).
\end{align}
\end{subequations}
Using these transformations and the phase factors for the chiral classification of Eq.~\eqref{eq: U(1) chiral PSG classification}, we find that the gauge fields on different bonds of the diamond lattice are given by
\begin{subequations}
\begin{align}
\overline{A}_{(0,0,0)_{A},(0,0,0)_{B}} &= \overline{A}  \\
\overline{A}_{(r_{1},r_{2},r_{3})_{A},(r_{1},r_{2},r_{3})_{B}} &= \overline{A}  \\
\overline{A}_{(r_{1},r_{2},r_{3})_{A},(r_{1}+1,r_{2},r_{3})_{B}} &= \overline{A} + \frac{n_{1/2} \pi}{2} (r_{3} - r_{2}) \\
\overline{A}_{(r_{1},r_{2},r_{3})_{A},(r_{1},r_{2}+1,r_{3})_{B}} &= \overline{A} - \frac{n_{1/2} \pi}{2} r_{3} \\
\overline{A}_{(r_{1},r_{2},r_{3})_{A},(r_{1},r_{2},r_{3}+1)_{B}} &= \overline{A} .
\end{align}
\end{subequations}
Fixing $\overline{A}=0$, we get the unit cell illustrated in Fig.~\ref{fig: Unit cell gauge field configurations}.

\section{\label{appendix: Constructing the MF Hamiltonian} Constructing the saddle point action}

To write down the GMFT action at the saddle point, we define the Fourier transform of the spinon field operator as
\begin{align}
    \Phi_{\mathbf{r}_{\alpha}}^{\tau} &= \frac{1}{\sqrt{\beta N_{\text{u.c.}}}}\sum_{i\omega_n} \sum_{\mathbf{k}} \Phi_{\mathbf{k},i\omega_n,r_{s},\alpha}e^{-i\left(\omega_n \tau - \mathbf{k}\cdot\mathbf{r}_{\alpha} \right)},
\end{align}
where $N_{\text{u.c.}}$ is the number of unit cells, $\beta=1/k_{B} T$ is the inverse temperature, and the position on the diamond lattice is
\begin{align}
    \mathbf{r}_{\alpha} &= \mathbf{r}_{\text{u.c.}} + \mathbf{r}_{s} - \frac{\eta_{\alpha}}{2}\mathbf{b}_{0}
\end{align}
with $\mathbf{r}_{\text{u.c.}}$ and $\mathbf{r}_s$ labeling the position of the GMFT \emph{Ansatz} unit cell and sublattice respectively. The wavevector sum is performed over the reduced first Brillouin zone associated with a GMFT \emph{Ansatz}. For the 0-, $\pi$- and $\pi/2$-flux, there are respectively 1, 4, and 16 sublattices per GMFT unit cell, as can be seen from Fig.~\ref{fig: Unit cell gauge field configurations}. To write down the action in of the $\pi$- and $\pi/2$-flux states in a compact form, we introduce the spinon field vector notation of Eq.~\eqref{eq: spinon vector field after FT} for both A and B diamond sublattices. After introducing these Fourier transformed vector fields, the GMFT action takes the general form given in Eq.~\eqref{eq: GMFT action after FT}. The spinon hopping matrix is defined by the relation
\begin{align}
    &\sum_{\mathbf{k},i\omega_n} \sum_{\alpha} \vec{\Phi}^{\dagger}_{\mathbf{k},i\omega_{n},\alpha} M^{\alpha}(\mathbf{k}) \vec{\Phi}_{\mathbf{k},i\omega_{n},\alpha} \nonumber \\
    &=-\frac{J_{\pm}}{4} \sum_{\mathbf{k},i\omega_n} \sum_{\mathbf{r}_{s},\alpha} \Phi_{\mathbf{k},i\omega_n,\mathbf{r}_s+\eta_{\alpha}\hat{\mathbf{e}}_{\mu},\alpha}^{*} \Phi_{\mathbf{k},i\omega_n,\mathbf{r}_s+\eta_{\alpha}\hat{\mathbf{e}}_{\nu},\alpha}    \nonumber \\
    &\exp\left[ -i \eta_{\alpha}  \left( \mathbf{k}\cdot\left( \hat{\mathbf{e}}_{\mu} - \hat{\mathbf{e}}_{\nu} \right) - \left( \overline{A}_{\mathbf{r_{s},\mathbf{r}_s + \eta_{\alpha} \mathbf{b}_{\nu}}} - \overline{A}_{\mathbf{r_{s},\mathbf{r}_s + \eta_{\alpha} \mathbf{b}_{\mu}}} \right) \right)  \right].
\end{align}

\section{\label{appendix: Evaluation of observables} Evaluation of observables}

\subsection{\label{appendix subsec: Evaluation of observables -> Green's function} Green's function}
The spinon Matsubara Green's function can be explicitly written by diagonalizing the spinon hopping matrix and inverting the right-hand side of Eq.~\eqref{eq: definition Matsubara Green's function} to get 
\begin{align}
    \mathscr{G}^{\alpha}_{\mu\nu}\left(\mathbf{k},i\omega_{n}\right) &= \expval{\vec{\Phi}_{\mathbf{k},i\omega_n,\mu,\alpha}^{\dagger} \vec{\Phi}_{\mathbf{k},i\omega_n,\nu,\alpha}} \nonumber \\
    &= \sum_{\gamma} \frac{2 J_{zz} U_{\nu\gamma}^{\alpha}(\mathbf{k}) U_{\gamma\mu}^{\alpha \dagger}(\mathbf{k}) }{\omega_n^{2} + 2J_{zz}\left(\lambda^{\alpha} + \varepsilon_{\gamma}^{\alpha}(\mathbf{k})\right)} ,
\end{align}
where the $U^{\alpha}(\mathbf{k})$ contains the eigenvector of $M^{\alpha}(\mathbf{k})$ and $\varepsilon^{\alpha}(\mathbf{k})$ are the corresponding eigenvalues (i.e., $U^{\alpha\dagger}(\mathbf{k})M^{\alpha}(\mathbf{k})U^{\alpha}(\mathbf{k}) = \text{diag}\left( \varepsilon^{\alpha}_{1}(\mathbf{k}), ..., \varepsilon^{\alpha}_{N_{\text{sl}}}(\mathbf{k})  \right)$). Performing an analytical continuation $i\omega_n\to \omega+ i\eta^{+}$ and identifying the poles of the retarded Green's function, we get the spinon dispersion $\mathcal{E}^{\alpha}_{\gamma}(\mathbf{k})=\sqrt{2J_{zz}(\lambda^{\alpha}+ \varepsilon_{\gamma}^{\alpha}(\mathbf{k}))}$. 

Performing the Matsubara sum and taking the $T\to 0$ limit yields 
\begin{align}
    \mathscr{G}^{\alpha}_{\mu\nu}\left(\mathbf{k}\right) &= \frac{1}{\beta}\sum_{i\omega_n}  \mathscr{G}^{\alpha}_{\mu\nu}\left(\mathbf{k},i\omega_{n}\right) \nonumber \\
    &= \sum_{\gamma} \frac{J_{zz} U_{\nu\gamma}^{\alpha}(\mathbf{k}) U_{\gamma\mu}^{\alpha \dagger}(\mathbf{k})}{\mathcal{E}^{\alpha}_{\gamma}(\mathbf{k})} \coth\left(\frac{\beta \mathcal{E}^{\alpha}_{\gamma}(\mathbf{k})}{2}  \right) \nonumber \\
    &\stackrel{T\to 0}{=} \sum_{\gamma} \frac{J_{zz} U_{\nu\gamma}^{\alpha}(\mathbf{k}) U_{\gamma\mu}^{\alpha \dagger}(\mathbf{k})}{\mathcal{E}^{\alpha}_{\gamma}(\mathbf{k})}.
\end{align}

\begin{figure}[t]
\includegraphics[width=1.0\linewidth]{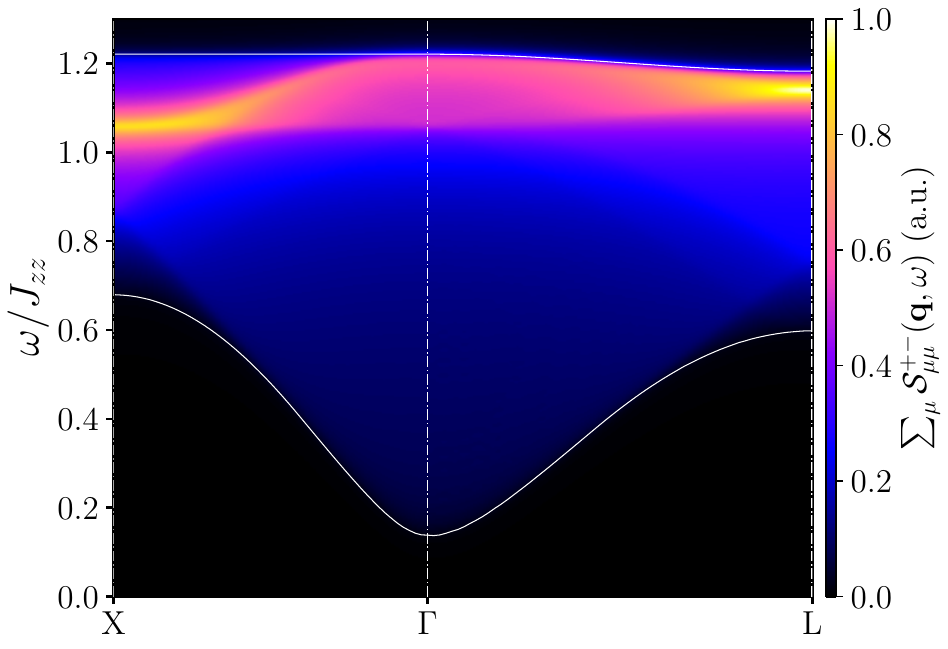}
\caption{Diagonal part of the dynamical spin structure factor in the local frame for the 0-flux state with $J_{\pm}/J_{zz}=0.046$. The results can be directly compared with the QMC calculations presented in Ref.~\cite{huang2018dynamics}.  \label{fig: comparison with QMC}}
\end{figure}

\subsection{\label{appendix subsec: Evaluation of observables -> Self-consistency condition} Self-consistency condition}

The sublattice-dependent Lagrange multiplier $\lambda^{\alpha}$ needs to be fixed such that the constraint
\begin{align}
    \frac{1}{N_{\text{d.u.c}}}\sum_{\mathbf{r}_{\alpha}}\expval{\Phi_{\mathbf{r}_\alpha}^{\dagger}\Phi_{\mathbf{r}_\alpha}} = \kappa
\end{align}
is respected for both $\alpha\in\left\{A,B\right\}$. Performing the sum and taking the $T\to 0$ limit leads to
\begin{align}
    \kappa &= \frac{1}{\beta N_{\text{d.u.c}}} \sum_{\mathbf{k}, i\omega_n} \sum_{\mu} \mathscr{G}^{\alpha}_{\mu\mu}(\mathbf{k},i\omega_n) \nonumber \\
    &\stackrel{T\to 0}{=} \frac{1}{N_{\text{d.u.c}}} \sum_{\mathbf{k}} \sum_{\gamma} \frac{J_{zz}}{\mathcal{E}^{\alpha}_{\gamma}(\mathbf{k})}.
\end{align}

\section{\label{appendix: Comparison of the dynamical spin structure factor for the 0-flux state with QMC} Comparison of the 0-flux state dynamical spin structure factor with quantum Monte Carlo}

We want to compare the dynamic correlations we obtain with GMFT for the 0-flux state with the QMC results presented in Ref.~\cite{huang2018dynamics}. In this QMC investigation, the sublattice-dependent dynamical correlations are defined as 
\begin{align}
\mathcal{S}^{+-}_{\mu \nu}(\mathbf{q},\omega)=&\frac{1}{N_{\text {u.c. }}} \sum_{\mathbf{R}_{\mu}, \mathbf{R}_{\nu}'} e^{i \mathbf{q} \cdot\left(\mathbf{R}_{\mu} - \mathbf{R}_{\nu}' \right)} \nonumber \\
&\hspace{1.2cm}\times\int \dd{t} e^{i \omega t}  \left\langle \mathrm{S}_{\mathbf{R}_{\mu}}^{+}(t) \mathrm{S}_{\mathbf{R}_{\nu}'}^{-}(0)\right\rangle,
\end{align}
where $\mathbf{R}_{\mu}$ and $\mathbf{R}_{\nu}$ label all sites of one of the four pyrochlore sublattices (i.e., $\mu,\nu\in\{ 0,1,2,3\}$) for the whole lattice and the spins are written in the local frame. The investigation reports the diagonal part of the dynamical spin structure factor $\sum_{\mu}\mathcal{S}^{+-}_{\mu\mu}(\mathbf{q},\omega)$ for the XXZ model with $J_{\pm}/J_{zz}=0.046$ along the $\Gamma\to\text{X}$ and $\Gamma\to\text{L}$ directions. To directly compare to these results, we compute the diagonal part of the dynamical spin structure factor for the 0-flux state with GMFT for the same coupling and along the same path in the first Brillouin zone. The results are presented in Fig.~\ref{fig: comparison with QMC}. These results are in surprisingly good agreement with QMC. First, the upper and lower bounds on the two-spinon continuum match. On top of being in the same energy range, subtle details like the flat upper edge of the two-spinon continuum along $\Gamma\to\text{X}$ in comparison to a minor decrease for the $\Gamma\to\text{L}$ path and a slightly lower position of the lower edge of the continuum at the L compared to the X points are captured. Next, the spectral weight behaves the same way. In both calculations, a broad continuum with most of the spectral weight close to the upper edge of the two-spinon continuum is observed. The spectral intensity increase along the paths $\Gamma\to\text{L}$ and $\Gamma\to\text{X}$ with maxima at the X and L points reported in QMC is further reproduced within GMFT. This surprising correspondence with QMC results should be taken as a compelling testimony to the reliability of GMFT. We again stress that this similarity is only possible with $\kappa=2$ for the rotor length self-consistency equation. With the usual choice of $\kappa=1$, the position of the two-spinon continuum is approximately twice as large, in complete disagreement with QMC.

%

\end{document}